\documentclass[final,3p,times,twocolumn]{elsarticle}

\usepackage{dsfont}
\usepackage{graphicx} 
\usepackage{graphics}

\usepackage{mathtools}
\usepackage{dcolumn}    
\usepackage{bm} 
\usepackage{graphicx}
\usepackage{amsmath}    
\usepackage{latexsym}
\usepackage{amsfonts}   
\usepackage{amssymb}
\usepackage{array}      
\usepackage{epsfig}
\usepackage{txfonts}
\usepackage{xcolor}
\usepackage{hyperref}
\usepackage{ulem}

\newcommand{\Tr}{\mbox{Tr}}

\newcommand{\ket}[1]{\left|#1\right\rangle}

\newcommand{\ketbra}[2]{|#1\rangle\langle#2|}

\journal{Control Engineering Practice}

\begin{document}

\begin{frontmatter}

\title{Suppressing leakage and maintaining robustness in transmon qubits: \\ Signatures of a trade-off relation}

\author{Pablo M. Poggi}
%\email{pablo.poggi@strath.ac.uk }
\ead{pablo.poggi@strath.ac.uk }
\affiliation{Department of Physics, SUPA and University of Strathclyde, Glasgow G4 0NG, United Kingdom }
\affiliation{Center for Quantum Information and Control, Department of Physics and Astronomy, University of New Mexico, Albuquerque, New Mexico 87131, USA}
\author{Anthony Kiely}
\affiliation{School of Physics, University College Cork, College Road, Cork, Ireland}
% \address{School of Physics, University College Dublin, Belfield Dublin 4, Ireland}
% \affiliation{Centre for Quantum Engineering, Science, and Technology,
% University College Dublin, Belfield, Dublin 4, Ireland}

\begin{abstract}
We study the problem of optimally generating quantum gates in a logical subspace embedded in a larger Hilbert space, where the dynamics is also affected by unknown static imperfections. This general problem is widespread across various emergent quantum technology architectures. We derive the fidelity susceptibility in the computational subspace as a measure of robustness to perturbations, and define a cost function that quantifies leakage out of the subspace. We tackle both effects using a two-stage optimization where two cost functions are minimized in series. Specifically, we apply this framework to the generation of single-qubit gates in a superconducting transmon system, and find high-fidelity solutions robust to detuning and amplitude errors across various parameter regimes. We also show control pulses which maximize fidelity while minimizing leakage at all times during the evolution. However, finding control solutions that address both effects simultaneously is shown to be much more challenging, indicating the presence of a trade-off relation.
\end{abstract}

%\maketitle

\begin{keyword}
Quantum control \sep Robust open loop control \sep Superconducting qubits
%% keywords here, in the form: keyword \sep keyword

%% PACS codes here, in the form: \PACS code \sep code

%% MSC codes here, in the form: \MSC code \sep code
%% or \MSC[2008] code \sep code (2000 is the default)

\end{keyword}

\end{frontmatter}

\section{Introduction} \label{sec:intro}

In the vast majority of quantum computing architectures, the fundamental building blocks (qubits) are naturally embedded in multi-level quantum systems, such as atoms or oscillators. Given the high fidelity requirements of quantum computing this sets two immediate criteria for successful quantum gates. Firstly, that the state evolution remains within the subspace of interest (often termed the logical or computational space) throughout the whole evolution. This will avoid the higher decoherence rates associated with these higher levels \cite{Wang2025_transmon} and other unknown errors. Secondly that the particular control pulses implement an evolution which is also robust to systematic errors within the logical space. This can be modeled by unknown static perturbations \cite{ruschhaupt2012,Poggi2024,WEIDNER2025}, which captures both systematic errors and quasi-static noise (fluctuations which are slower than the system timescales, common in low-frequency dominated noise) \cite{bylander2011}.

Previously, these two issues have been addressed separately. Leakage errors have been combated using shorcuts to adiabaticity \cite{guery2019,kiely2014}, Derivative Removal by Adiabatic Gate (DRAG) \cite{motzoi2009_drag,babu2021}, composite pulses \cite{genov2013,cykiert2024,tonchev2025} and numerical optimization \cite{rebentrost2009,safaei2009,werninghaus2021,Hyyppa2024}. Similarly, a variety of methods have been employed to ensure robust quantum control such as geometric curves \cite{Dong2021}, numerical optimization using differentiable programming \cite{Coopmans2021,Coopmans2022} and gradient free methods \cite{turyansky2025}, Pontryagin maximum principle \cite{Fresse-Colson2025} and many others \cite{khodjasteh2012, Daems2013,Araki2023,Irtaza2023,Propson2022}. In particular, in \cite{Poggi2024} it was recently shown how control fields can be designed which correct for {\it any} unknown static error. In both cases, an increase in operation time was required.

Our research question now is the following: {\it Can these effects be mitigated simultaneously, while still achieving high error-free gate fidelities?} Enforcing the evolution to avoid any leakage necessarily reduces the solution space and constrains the possible evolution paths. Since the fidelity susceptibility to static errors (i.e. the robustness) has been shown to be highly path dependent, heuristically it would seem that these demands could be in conflict.

Superconducting transmon qubits are an ideal testbed to explore this interplay. They are weakly anharmonic oscillators where the two lowest levels form the qubit. Although this parameter regime of weak anharmonicity allows for resistance to charge noise, the leakage to higher levels must be carefully managed. Previously, transmon control pulses have been successfully obtained using quantum optimal control approaches \cite{heeres2017,abdelhafez2020,gautier2025}, and more recently reinforcement learning methods \cite{nguyen2024}. Note that we will focus exclusively on coherent errors. However, incoherent errors are also present in superconducting systems \cite{papivc2023,tripathi2024}. The errors modeled by Lindblad dynamics can be partially mitigated by a suitable choice of control function \cite{levy2018}.

Our agenda for this paper is to investigate numerically optimized high-fidelity quantum gates which minimize errors due to both leakage and static perturbations. Our contributions are three-fold. First, we derive a new closed-form expression for the fidelity susceptibility of a logical operation within a computational space embedded in a larger Hilbert space. While we focus on its use for the example of a transmon qubit, this expression is general and is directly applicable to other platforms. Then, we propose and implement a two-stage optimization procedure (inspired by previous work \cite{kosut2022,Poggi2024}) that allows us to systematically investigate how demanding increased robustness or reduced leakage to the  optimal control solutions affect their cost in terms of increased evolution time or anharmonicity. Finally, we numerically reveal an emerging trade-off between the robustness and the leakage of the obtained control solutions and explore how it can be understood and mitigated. 
 
The paper is organized as follows. In Section \ref{sec:robust_qc}, we introduce the problem of generating robust quantum gates. We define the cost functions and derive the fidelity susceptibility to quantify robustness. We then present our two-stage optimization method. In Section \ref{sec:transmon}, we apply this framework to a transmon qubit. We show in Section \ref{sec:results_robust} that we can generate highly robust single-qubit gates and analyze the control resources required. In Section \ref{sec:leakage_tradeoff}, we examine leakage as an alternative optimization metric, and discuss the aforementioned trade-off between leakage and robustness. Finally, in Section \ref{sec:conclusions}, we summarize our findings and discuss future research directions.

\begin{figure}[t!]
    \centering    \includegraphics[width=\linewidth]{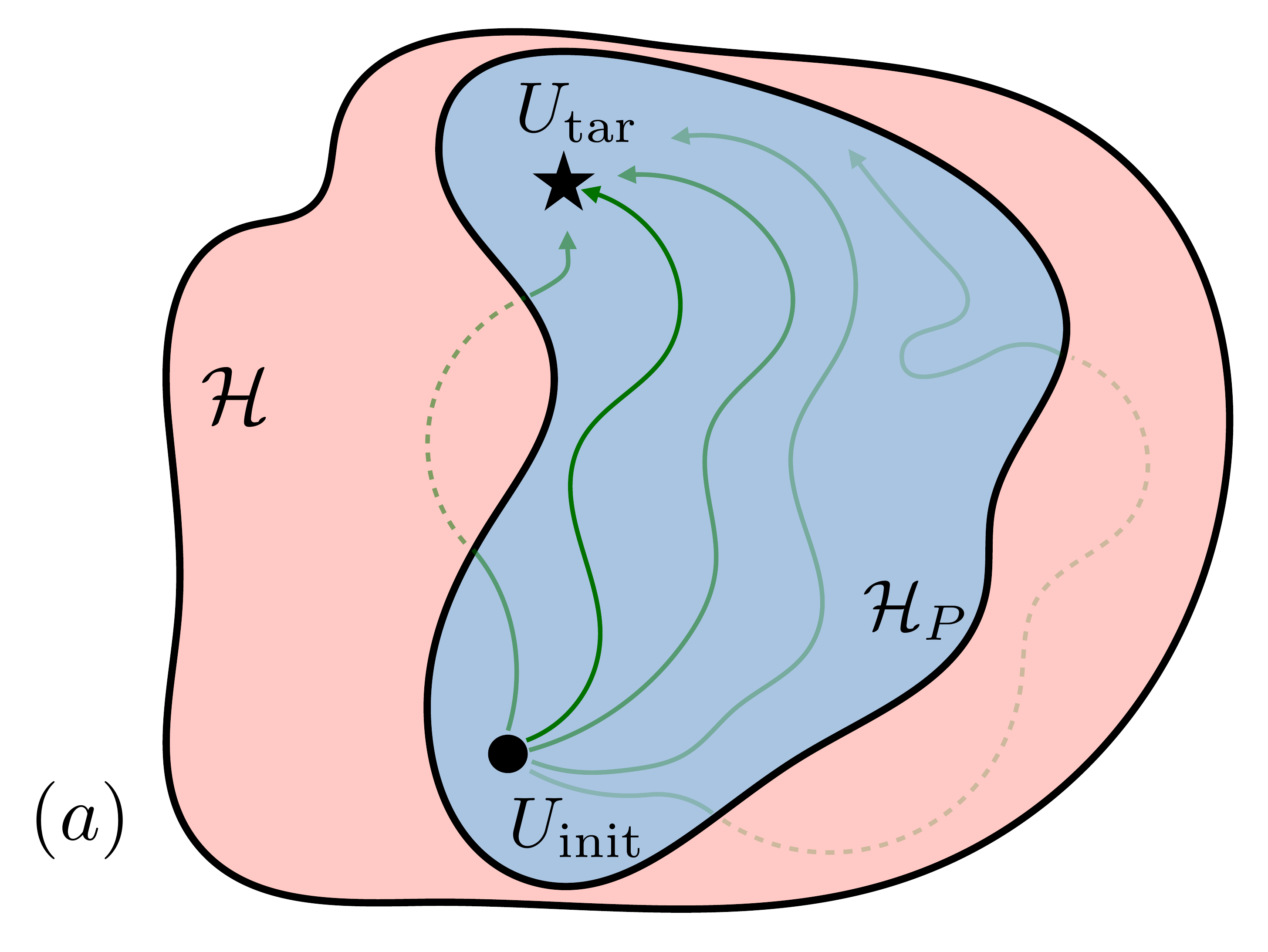}
    \includegraphics[width=\linewidth]{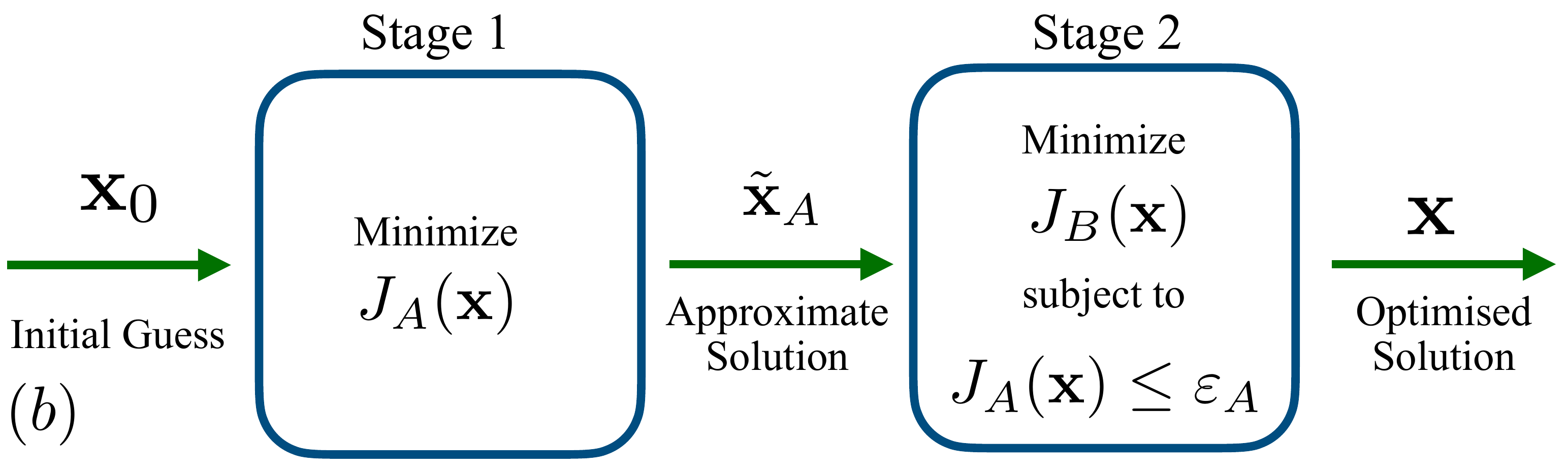}
    \caption{Outline of the relevant control problem.
    (a) Diagram of possible evolutions of the system (green lines) from the initial unitary $U_{\rm init}=1$ to the target one $U_{\rm tar}$. (b) Workflow of two-stage optimization. Inputs and output control pulse vectors show by green arrows and optimisation stages in blue boxes.}
    \label{fig:fig1}
\end{figure}

\section{Robust quantum control in a subspace} \label{sec:robust_qc}

We will now outline the problem setting in detail. Consider a general driven quantum system, described by an ideal, error-free Hamiltonian $H_0(t)$. The associated unitary time evolution operator from time $0$ to time $t$ is $U_0(t,0)\equiv U_0(t)$, which acts on states on a Hilbert space $\mathcal{H}$. We will consider a partition of whole space into a relevant computational subspace $\mathcal{H}_P$ and its complement $\mathcal{H}_{\overline{P}}$, i.e. $\mathcal{H}=\mathcal{H}_P\oplus \mathcal{H}_{\overline{P}}$, see Fig. \ref{fig:fig1} (a) for an illustration. We are interested in transformations only within this computational subspace; in particular we want $U_0(t)$ to be such that a the final time $t=T$, any state $\ket{\psi}\in \mathcal{H}_P$ transforms to $U_{\rm tar}\ket{\psi}$ up to an irrelevant global phase. Any unitary obeying
\begin{equation}
    U_0(T) = U_{\rm tar} \oplus U_{\overline{P},T}
\end{equation}
(up to a global phase) will satisfy this criterion. 

To quantify how well our dynamics achieves this goal, we resort to the state-averaged fidelity, originally derived in \cite{pedersen2007fidelity}. We will focus on the application of this formula to a relevant computational subspace, see for example \cite{fromonteil2023protocols}. As such, we define the fidelity in the logical space between two unitaries $X$ and $Y$ as,
\begin{align}
    &G[X,Y]=\nonumber \\ &\frac{1}{d_P(d_P+1)} \left\{\Tr[X P X^\dagger P Y P Y^\dagger P]+ \left|\Tr[P X P Y^\dagger]\right|^2 \right\}, 
    \label{eq:general_fidelity}
\end{align}
where $P$ is a projection operator into the subspace of interest, $d_P=\rm{dim}(\mathcal{H}_P)$ is the dimension of this subspace. In this language, we can define a cost function that our dynamics should seek to minimize in order to achieve the desired target. This target cost reads
\begin{eqnarray}
    J_U= 1- G[U_{\rm tar} \oplus 1, U_0(T,0)].\label{eq:cost_tar}
\end{eqnarray}
Minimizing this will ensure that the target gate is implemented by the ideal Hamiltonian $H_0(t)$.

We now consider two types of imperfections that could affect our idealized control solutions. Firstly, there could be static perturbations to this Hamiltonian, resulting in the full Hamiltonian $H(t)= H_0(t) + \lambda V$ where $\lambda \ll 1$. To quantify the deviation of $U_\lambda$ with respect to $U_0$ we can use Eq. (\ref{eq:general_fidelity}) with $X=U_0$ and $Y=U_\lambda$ to obtain
\begin{equation}
    F_\lambda = \frac{1}{d_P(d_P+1)} \left\{\Tr[P U_\lambda P U_\lambda^\dagger]+ \left|\Tr[P U_\lambda P U_0^\dagger]\right|^2 \right\} , \label{eq:fidP}
\end{equation}
where the time evolution operator $U_\lambda$ is taken the final time $U_\lambda=U_\lambda(T)$ and we have assumed that the ideal evolution at the final time $U_0=U_0(T)$ has no leakage.

Robustness to a small perturbation $\lambda$ can then be characterized by the fidelity susceptibility derived in \ref{app:Fid_sus} as,
\begin{align}
\frac{d^2 F_\lambda}{d \lambda^2} \Big|_{\lambda=0} &=  - \frac{2 T^2}{d_P}  \Big\{ \Tr_P[\bar{V}_0^2]\nonumber \\
  &  -\frac{1}{d_P+1}\left[ \Tr_P[\bar{V}_0]^2+\Tr_P[\bar{V}_0 P \bar{V}_0]\right] \Big\},
\end{align}
where we have defined $\Tr_P[\cdot]=\Tr[P \cdot]$ as the trace over the logical subspace and have introduced the time-averaged perturbation operator
\begin{equation}
    \bar{V}_0 = \frac{1}{T}\int_0^T  U_0(t)^\dagger V U_0(t) dt.
\end{equation}
This will form the basis for the second cost function
\begin{equation}
    J_R = -\frac{1}{2\Omega^2T^2}\frac{d^2 F_\lambda}{d \lambda^2} \Big|_{\lambda=0}.
    \label{eq:cost_robust}
\end{equation}
\noindent where $\Omega$ is a characteristic frequency chosen to keep $J_R$ dimensionless (this will be specified in Sec. \ref{sec:transmon}). Note that evaluation of this cost function requires only information about the ideal (perturbation-free) evolution $U_0(t)$, depending only on $H_0(t)$.

Secondly, we will analyze loss of population outside of the computational subspace during the evolution. We quantify this using the expression
\begin{equation}
    L[U] = 1 - \frac{1}{d_P}\Tr\left(PUPU^\dagger\right),
    \label{eq:leakage_avg}
\end{equation}
\noindent which measures how population is lost when a state is evolved by $U$, averaged over all initial states contained in $P$. Consequently, $L[U]=0$ only if $U=U_P\oplus U_{\overline{P}}$. In order to find solutions $U_0(t)$ that minimize leakage throughout the evolution, we thus define the final cost functional
\begin{equation}
    J_L = \frac{1}{T}\int_{0}^T L[U_0(t)] \,dt.\label{eq:cost_leak}
\end{equation}

We emphasize that reducing leakage at the \textit{final} time is already achieved by the target cost function $J_U$, c.f. Eq. (\ref{eq:cost_tar}). Considering this additional cost function $J_L$ aims to keep the population in the computational subspace throughout the complete dynamics.

Fig. \ref{fig:fig1}(a) shows a schematic of the control task, where the three functionals have a clear meaning. $J_U$ measures how close the final unperturbed unitary $U_0(T)$ matches the target (marked by the star). $J_L$ quantifies the leakage (dashed green lines) outside the computational Hilbert space $\mathcal{H}_P$ (shaded blue area). Finally, $J_R$ captures the deviations in gate fidelity due to variation in the error parameter $\lambda$. The increasing transparency of lines corresponds to stronger perturbations. Note that the perturbations $V$ reside in the entire Hilbert space and could therefore induce transitions outside the computational subspace, in addition to reducing overall fidelity.

\subsection{Optimization procedure}
Having defined the cost functions of interest, i.e. Eqns. \eqref{eq:cost_tar}, \eqref{eq:cost_robust}, and \eqref{eq:cost_leak}, we now discuss how to tackle the multi-objective optimization procedure in practice. We will focus on bicriteria problems where two functionals $J_A(\mathbf{x})$ and $J_B(\mathbf{x})$ are meant to be jointly optimized and $\mathbf{x}$ is a vector which parametrizes the control fields. A simple approach to achieve this is to define a unified scalar cost function as $a J_A + b J_B$ with $a,b>0$. Since $J_A,J_B\geq 0$, the global minimum of $J$ is achieved if and only if the global minima of $J_A$ and $J_B$ are achieved separately (\cite{miettinen1999}, see also \cite{shao2024}). However, this introduces new variables $a,b$ into the problem which often need to be varied ad hoc. 

Here we consider a two-stage optimization procedure inspired by Ref. \cite{kosut2022,kosut2023} (see also \cite{mavrotas2009}) and already applied to a robust optimal control problem in \cite{Poggi2024}. In the first stage of the optimization, we seek to minimize $J_A(\mathbf{x})$ alone starting from a (typically random) initial guess $\mathbf{x}_0$. Once a suitable approximate solution $\tilde{\mathbf{x}}_A$ has been found, we feed this as the initial guess for stage two. Here, we seek to minimize $J_B(\mathbf{x})$ but under the constraint that 
\begin{equation}
    J_A(\mathbf{x})\leq \varepsilon_A
\end{equation}
i.e., that the solution does not deviate too far away from the optimal manifold of the first stage. The workflow for this two-stage optimization approach is illustrated in Fig. \ref{fig:fig1}(b). In Sec. \ref{sec:results_robust} we will consider a combined target \& robustness optimization where $J_A=J_U$ and $J_B=J_R$. Then, in Sec. \ref{sec:leakage_tradeoff} we will study a target \& leakage search, where $J_A=J_U$ and $J_B=J_L$, and explore the extension to multi-objective problems where all three cost functions $J_U$, $J_R$, and $J_L$ are minimized. 

In practice, we can tackle the first stage with standard methods found in most off-the-shelf optimization libraries. In our case we use the L-BFGS-B method \cite{zhu1997,byrd1995} of \texttt{scipy.optimize} in Python. The second stage requires the implementation of a non-linear constraint on the search space, a feature which is not compatible with every standard optimization routine. However, we find that such constraints can be easily introduced in the Sequential Least Squares Programming (SLSQP) implementation of the same library. We find good results using this combination of methodologies.

\section{Application to the Transmon system} \label{sec:transmon}

Superconducting transmon qubits \cite{devoret2004,Gao2021} stand out as a highly promising platform for quantum computing due to their scalability and fast gate times. A single transmon system can be modeled as a driven anharmonic oscillator, where the qubit is encoded in the the lowest two levels. Thus, in this setting the relevant projector into the logical subspace is $P=\ketbra{0}{0}+\ketbra{1}{1}$ and $d_P=2$.

Setting the frequency of the drive $\omega_d$ detuned from the frequency $\omega_{01}$ of the first transition of the oscillator by $\delta$, and applying the rotating wave approximation yields the Hamiltonian \cite{nguyen2024}

\begin{equation}
    H_0(t) = \left(\delta-\frac{\alpha}{2}\right)\hat{n} + \frac{\alpha}{2}\hat{n}^2 + \frac{\Omega}{\sqrt{2}}\left[d_R(t) \hat{q} - d_I(t) \hat{p}\right].
\end{equation}
The number operator $\hat{n}=a^\dagger a$ is defined in terms of the usual bosonic annihilation operator $a$. Similarly, we have $\hat{p}=i(a^\dagger-a)/\sqrt{2}$ and $\hat{q}=(a+a^\dagger)/\sqrt{2}$.

The relevant parameters of the model are the detuning $\delta = \omega_{01}-\omega_d$ and the anharmonicity $\alpha = \omega_{12}-\omega_{01}$, where $\omega_{ba}$ is the frequency difference between the bare states $\ket{b}$ and $\ket{a}$.  The control is parametrized by two independent dimensionless fields $\{d_R(t), d_I(t)\}$. Unless stated otherwise we take $\alpha/\Omega=-2$, $\delta/\Omega=-0.5$, and measure evolution times in units of $T_{\Omega}=2\pi/\Omega$. For our numerical optimizations, we find that truncating Hilbert space at $N=6$ levels is enough to balance convergence and computational cost.

As discussed in Sec. \ref{sec:robust_qc}, we are interested in the problem of realizing a target gate in the computational subspace $U_{\mathrm{tar}}$ robustly in the presence of an unknown weak perturbation $\lambda V$. Here we will consider the case of $V=\hat{n}$ (detuning errors), $V=\hat{q}$ (amplitude errors) and $V=\hat{n}^2$ (anharmonicity errors). To ensure a fair comparison, we rescale $\lambda$ by $\Tr_P(V^2)$ which is 1 for $V=\hat{n}$ and $V=\hat{n}^2$, and 2 for $V=\hat{q}$. We will focus on a fixed target transformation $U_{\mathrm{tar}}=X=\ketbra{0}{1}+\ketbra{1}{0}$, but naturally our scheme is general and applies to arbitrary targets.

For our numerical calculations, We choose a piecewise-constant parametrization where both fields $\{d_R(t),d_I(t)\}$ take constant values $\{d_R^{(j)},d_I^{(j)}\}$ in each time-interval $(j-1)\Delta t \leq t < j\Delta t$, where $j=1,\ldots,M$ and $M$ is the number of time intervals. In our numerical optimization we constrain $|d_R^{(j)}|,\: |d_I^{(j)}|\leq 1$ and fix $M=15$ irrespective of the choice of evolution time $T$. This choice balances the expressivity of our ansatz with the numerical optimization cost. Our conclusions do not change significantly if $M$ is varied, although dropping $M$ below 10 starts showing signs of an overly constrained optimization problem. 

In the limit of large anharmonicity $\alpha$, this system can be treated as an actual qubit and the desired $X$ gate can be implemented by a standard $\pi$-pulse (i.e., setting $d_R=1$, $d_I=0$, $\delta=0$ for $T=T_{\Omega}/2$). However, the desire for high-precision operations require us to consider the effects of finite $\alpha$. A well-established way to do this is to consider the DRAG protocol \cite{motzoi2009_drag}, which leverages control over the off-quadrature driving $d_I(t)$ and detuning $\delta(t)$ to suppress leakage outside the computational subspace. This approach has been successfully demonstrated experimentally \cite{Hyyppa2024,wang2025}. Different variants of DRAG exist \cite{theis2018}; in the next section we will compare the performance of our robust optimal control solutions to the optimal first-order DRAG \cite{gambetta2011} which we describe in detail in \ref{app:DRAG}.

\section{Robust single qubit gates} \label{sec:results_robust}

\begin{figure} 
    \centering
    \includegraphics[width=\linewidth]{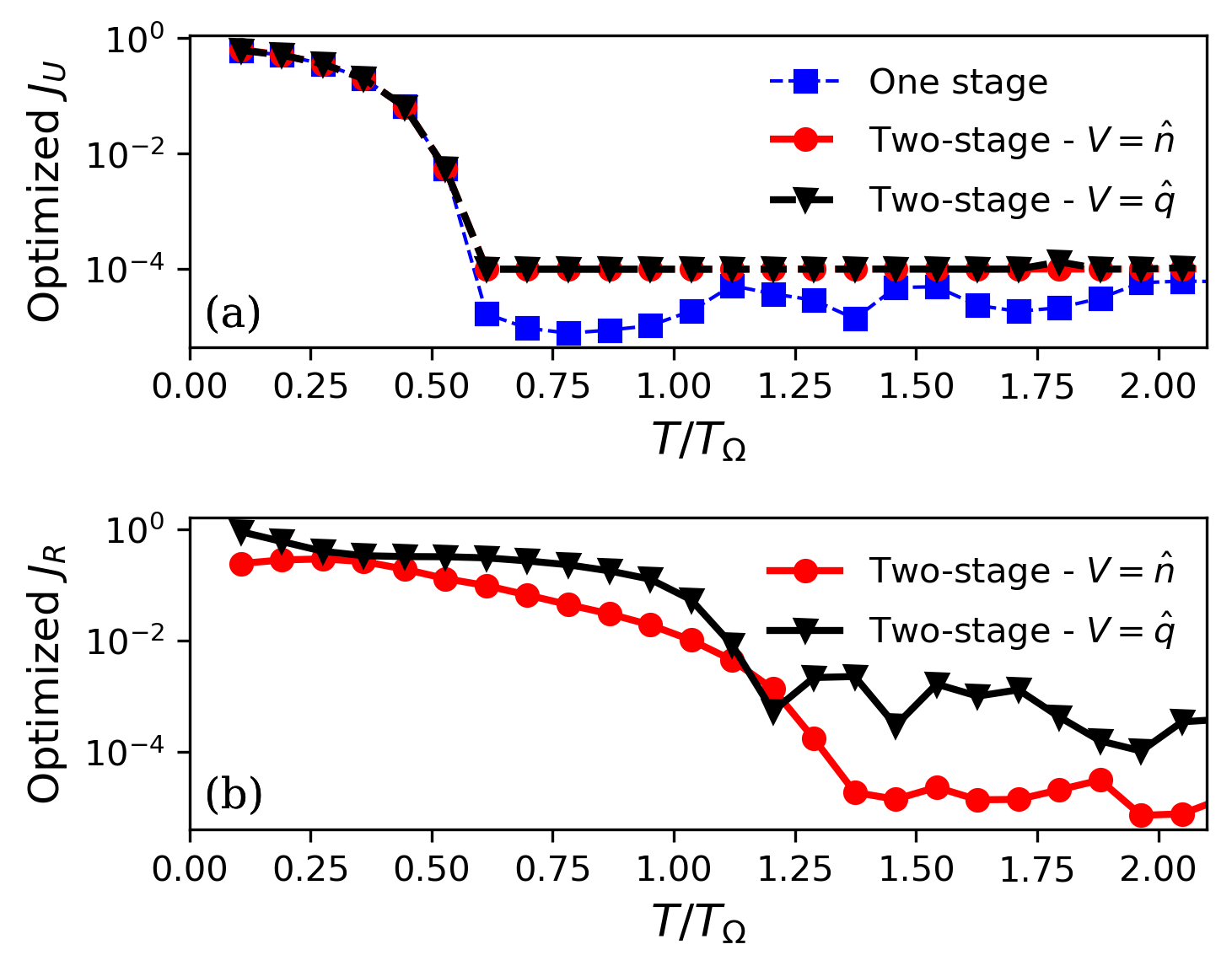}
    \caption{Optimized cost functions (a) target $J_U$ and (b) robustness $J_R$ against gate time $T$ for single stage (blue) and two-stage (black/red) optimization. Red dots correspond to choosing $V=\hat{n}$ as the perturbation operator, while black triangles to $V=\hat{q}$. For two-stage optimizations we have fixed the threshold at $\varepsilon_U = 10^{-4}$. Other parameters are set at $\alpha/\Omega=-2$, $\delta/\Omega=-0.5$.}
    \label{fig:control_time}
\end{figure}

We begin our analysis by identifying the minimum evolution time \cite{deffner2017,poggi2019} required to implement the desired target gate given the parameters of the problem. To do this we systematically run the optimization procedure for various choices of total evolution time in the range $T/T_{\Omega}\in[0,2]$. We compare the performance of the optimization in three different cases: a target-only search (blue squares), and two instances of the two-stage target \& robust search, corresponding to different choices of the perturbation operator (red circles for $V=\hat{n}$ and black triangles for $V=\hat{q}$). To sweep the range of evolution times, we begin considering the largest value of $T$ and random initial guesses for $\{d_R^{(j)},d_I^{(j)}\}$ ($j=1,\ldots,M$). As $T$ is lowered, we use as an initial guess the optimal value of the previous optimization to warm-start the search (see \cite{duncan2025} for other examples of this approach).

Results are shown in Fig. \ref{fig:control_time}. For all cases considered, we display the value of the optimized target functional $J_U$ (after the last stage) and the robustness functional $J_R$ as a function of the total evolution time in the top (a) and bottom (b) panels, respectively. For the target functional $J_U$ we see a distinctive sharp increase as $T/T_\Omega$ becomes small. This behavior is typical in small-dimensional quantum systems \cite{caneva2009,poggi2019} and indicates a clear separation between a controllable and an uncontrollable phase of the control problem \cite{bukov2018}. In this case, this allows us to estimate that the minimum control time to achieve this target (and with the given choice of system parameters) is $T\simeq T_\Omega/2$, which coincides with the time required for a simple $\pi$-pulse. 

\begin{figure}
    \centering
    \includegraphics[width=\linewidth]{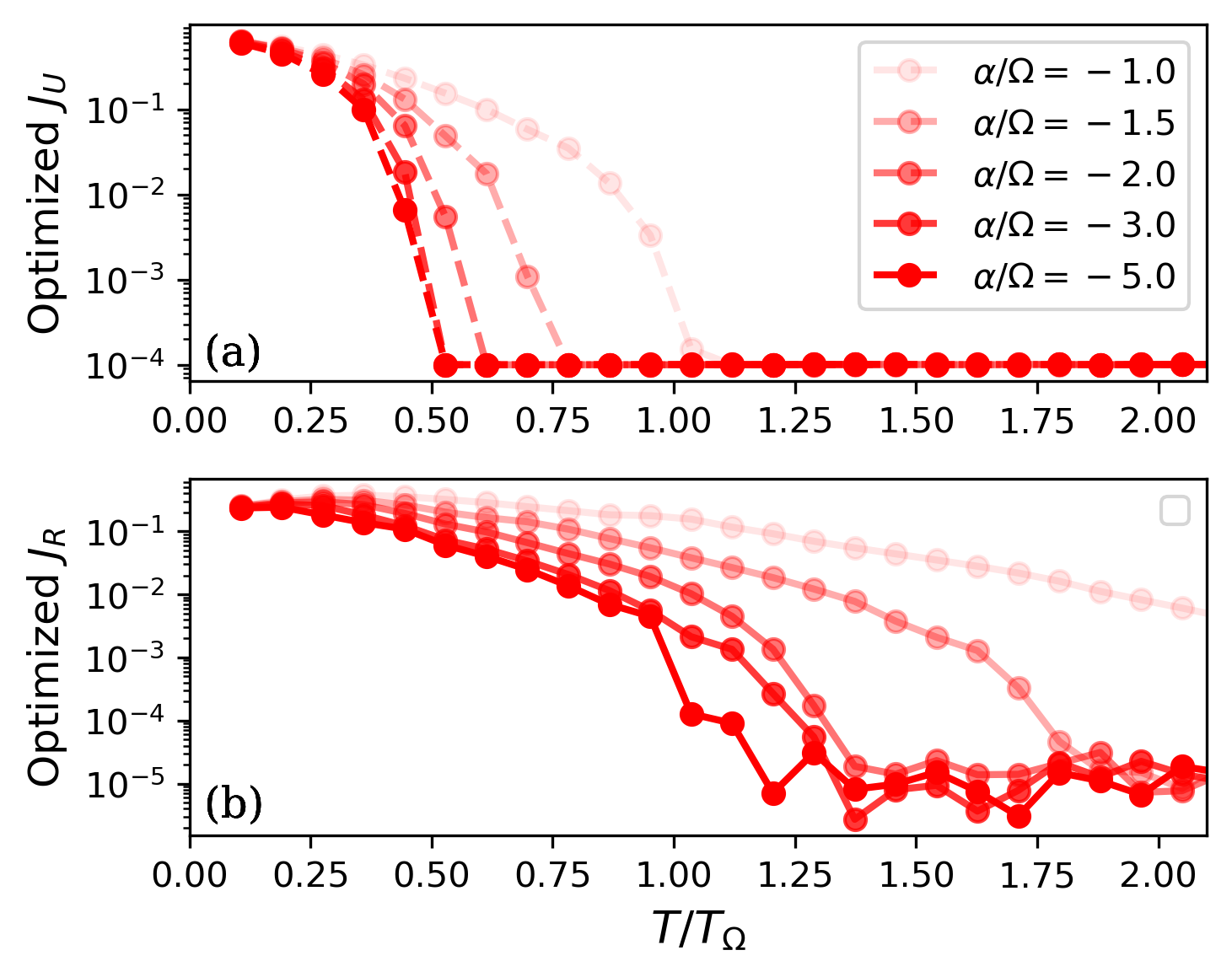}
    \caption{Optimized cost functions (a) target $J_U$ and (b) robustness $J_R$ against gate time $T$ for different value of anharmonicity $\alpha$. Results are shown for $V=\hat{n}$, $\delta/\Omega=-0.5$. In the optimization the threshold is set at $\varepsilon_U = 10^{-4}$.}
    \label{fig:control_time_alpha}
\end{figure}

Interestingly, the shape of the optimized target functional remains largely unchanged when we consider two-stage optimizations. As expected, however, we find that considerable more evolution time is required to achieve robustness to the studied perturbations. While the functional $J_R$ does not show such a steep change as $T$ is varied, a visual estimate indicates that an evolution time of $T\simeq 1.2 T_\Omega$ is required in order to achieve $J_R \sim 10^{-3},10^{-4}$. We corroborate this estimate numerically by analyzing different values of the anharmonicity parameter $\alpha$ for the case of $V=\hat{n}$. Results are shown in Fig. \ref{fig:control_time_alpha}, where we observe diminishing returns in both robustness and evolution time from increasing the anharmonicity $|\alpha/\Omega|$ above 2. From \ref{app:Leak} and similar bounds relating fidelity and leakage \cite{kiely2014}, one would expect that high fidelity gates require operation times far exceeding $1/|\alpha|$.

\begin{figure*}
    \centering
    \includegraphics[width=0.33\linewidth]{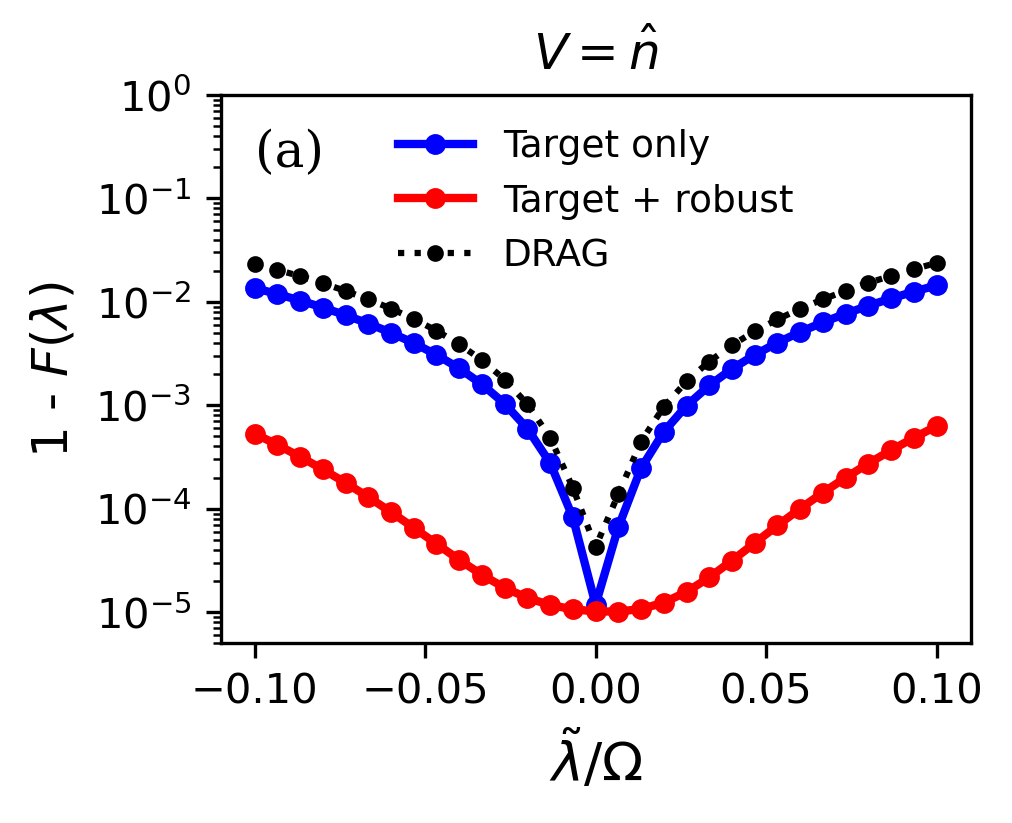}
    \includegraphics[width=0.33\linewidth]{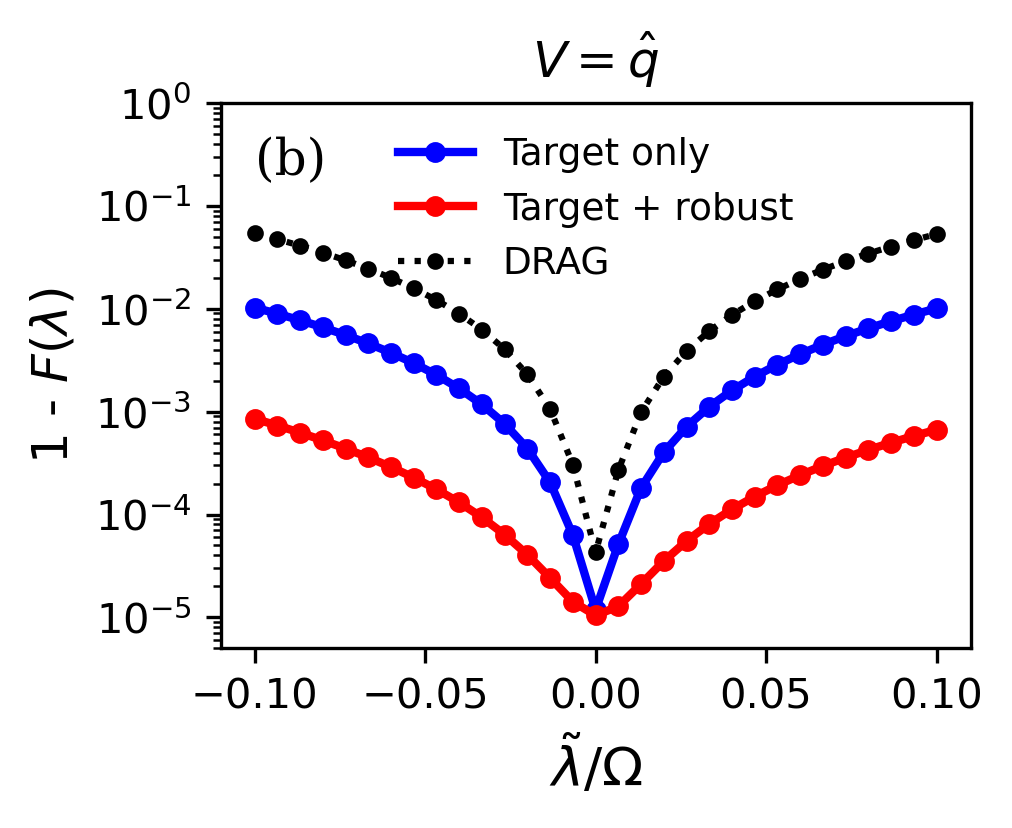}
    \includegraphics[width=0.33\linewidth]{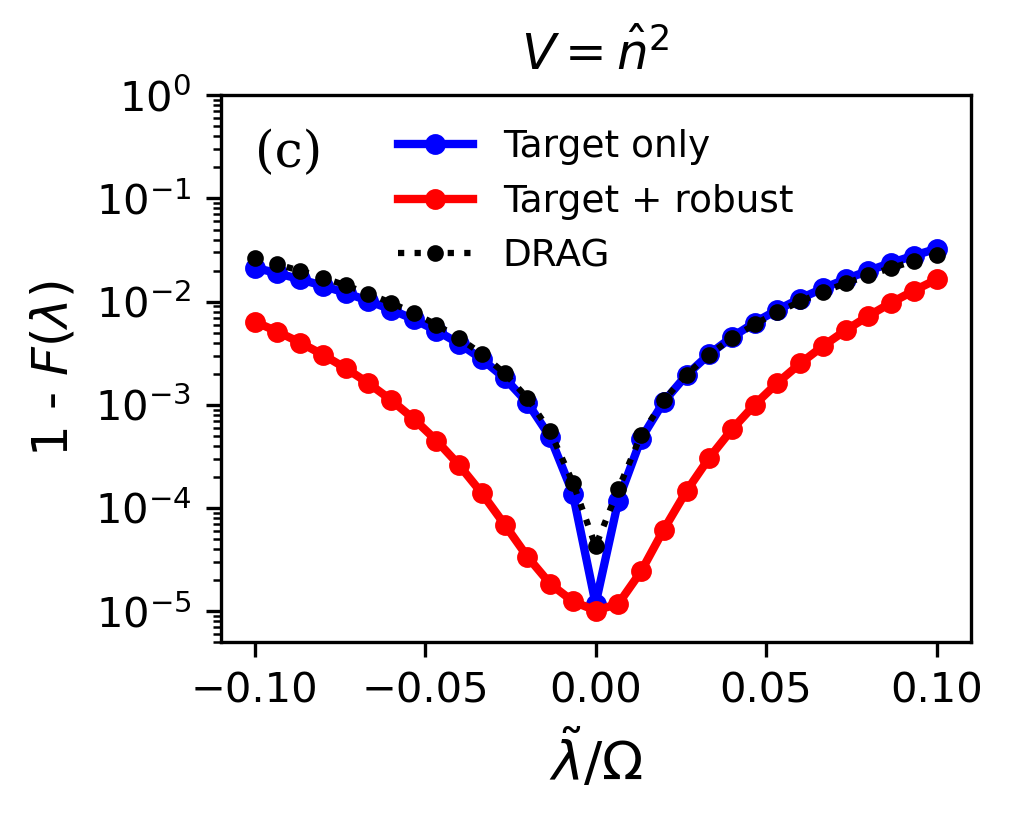}
    \caption{Average gate fidelity, as defined in Eq. (\ref{eq:fidP}), for various control protocols in presence of a static perturbation $\lambda V$. Each column corresponds to a different choice of $V$, and $\tilde{\lambda}=\lambda/\Tr_P(V^2)$ is the rescaled perturbation strength. Blue: target-only optimization with $T/T_{\Omega}=0.6$. Red: target \& robustness optimization with $T/T_{\Omega}=1.3$.  Threshold in two-stage optimization is fixed at $\varepsilon_U = 10^{-5}$. Black dotted lines show the performance of simple versions of DRAG pulses with evolution time $T/T_{\Omega}=1.3$, and setting $\sigma/T_{\Omega}=0.369$ (see \ref{app:DRAG} for the definition of these parameters; values were chosen to optimize the fidelity at zero perturbation). Other parameters were set to $\alpha/\Omega=-2$ and $\delta/\Omega=-0.5$ (for the optimal control simulations). }
    \label{fig:perturb_plots}
\end{figure*}

The fact that robust control solutions require more time than regular control solutions is natural given the additional constraint in the optimization problem. To study the gain obtained by investing extra control resources, we analyze and compare the performance of various control solutions, for fixed evolution times, in presence of a static perturbation. In Fig. \ref{fig:perturb_plots} we show the average infidelity as a function of the perturbation strength $\lambda$ for various cases of interest. The fidelity  is obtained by numerically simulating the dynamics of the system $H_0(t)+\lambda V$ and computing Eq. \eqref{eq:fidP} at the final time $T$ (recall $U_{\rm tar}=X$). To ensure the convergence of the optimal control solutions, these simulations are ran in a larger Hilbert space with $N=11$ levels. Each plot in the figure shows the comparison between the dynamics resulting from the target-only optimization (blue) at $T/T_{\Omega}=0.6$ and the dynamics coming from the target \& robustness optimization with $T/T_{\Omega}=1.3$ (red). Different columns in the figure correspond to different choices of perturbation operator, namely $V=\hat{n}$ in (a), $V=\hat{q}$ in (b), and $V=\hat{n}^2$ in (c). For completeness, we also include the performance of the DRAG protocol described in the previous section and in \ref{app:DRAG} (black), where we take $T/T_{\Omega}=1.3$ and $\sigma/T_{\Omega}=0.369$. These parameters were chosen to optimize the fidelity at $\lambda=0$ while still keeping the evolution time comparable to the other approaches.

The results in Fig. \ref{fig:perturb_plots} confirm that the optimal robust control solutions are indeed much less sensitive to the action of the perturbation than the other control protocols. Robust solutions are able to maintain average gate fidelities of at least $F\sim 1- 10^{-3}$ while allowing roughly $10\%$ deviations in amplitude and detuning, and at least $F\sim 1-10^{-2}$ for similar deviations in anharmonicity. The acquired robustness is evident from comparing the curvature of the infidelity plots at $\lambda=0$, which is exactly what the fidelity susceptibility c.f. Eq. (\ref{eq:cost_robust}) measures. These results show that `spending' the extra evolution time is indeed beneficial for the performance of the gate as it allows the system to reach better fidelities in presence of the static perturbations. 

\section{Leakage vs robustness trade-off} \label{sec:leakage_tradeoff}

\begin{figure}[h]
    \centering
    \includegraphics[width=\linewidth]{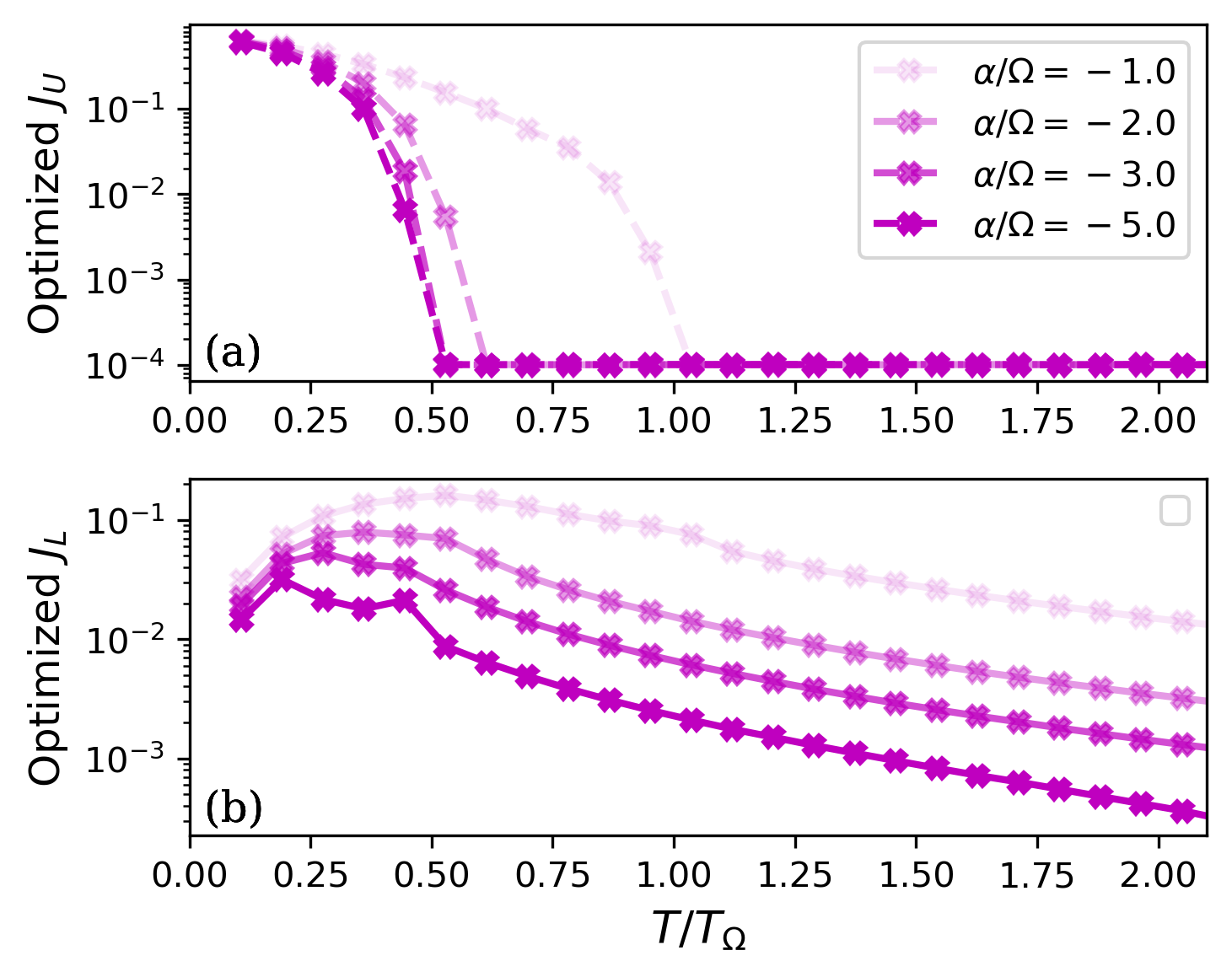}
    \caption{Optimized cost functions (a) target $J_U$ and (b) leakage $J_L$ against gate time $T$ the two-stage optimization target \& leakage. We have fixed the threshold at $\varepsilon_U = 10^{-4}$. Other parameters are set at $\alpha/\Omega=-2$, $\delta/\Omega=-0.5$.}
    \label{fig:control_time_leak}
\end{figure}

\begin{figure*}
    \centering
    \includegraphics[width=0.65\linewidth]{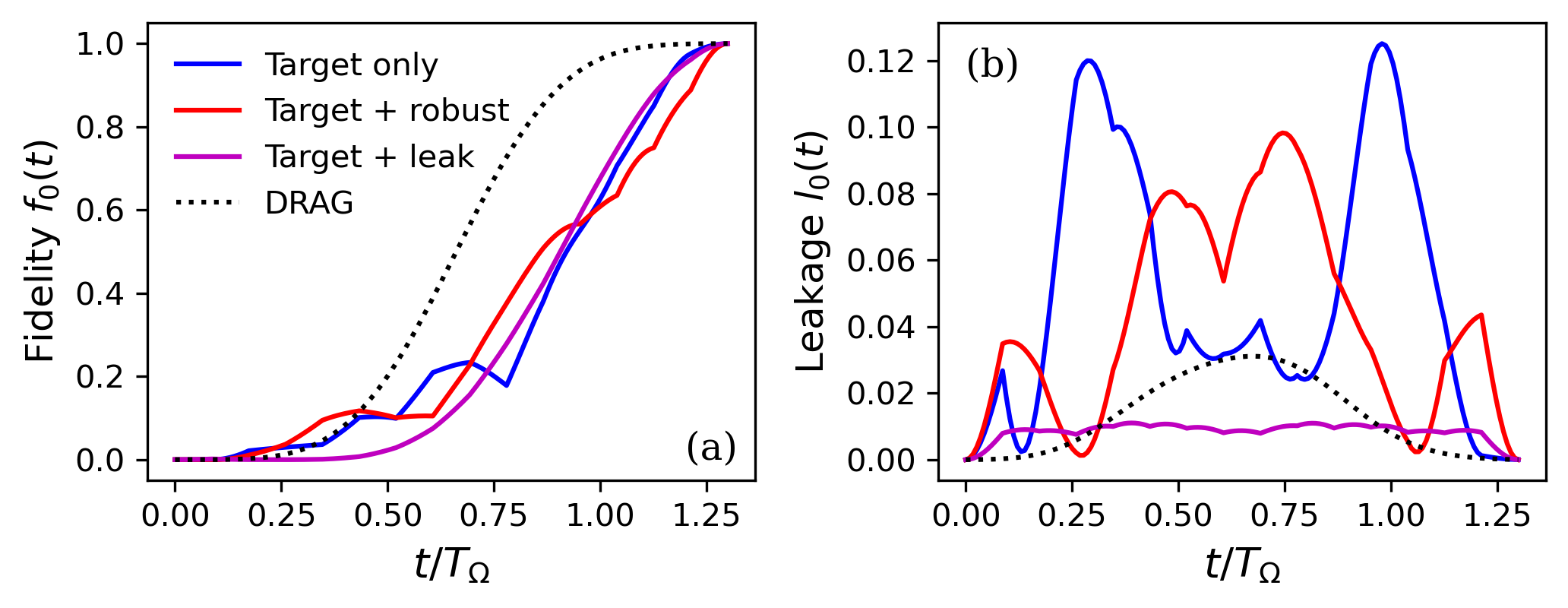}    
    \includegraphics[width=0.33\linewidth]{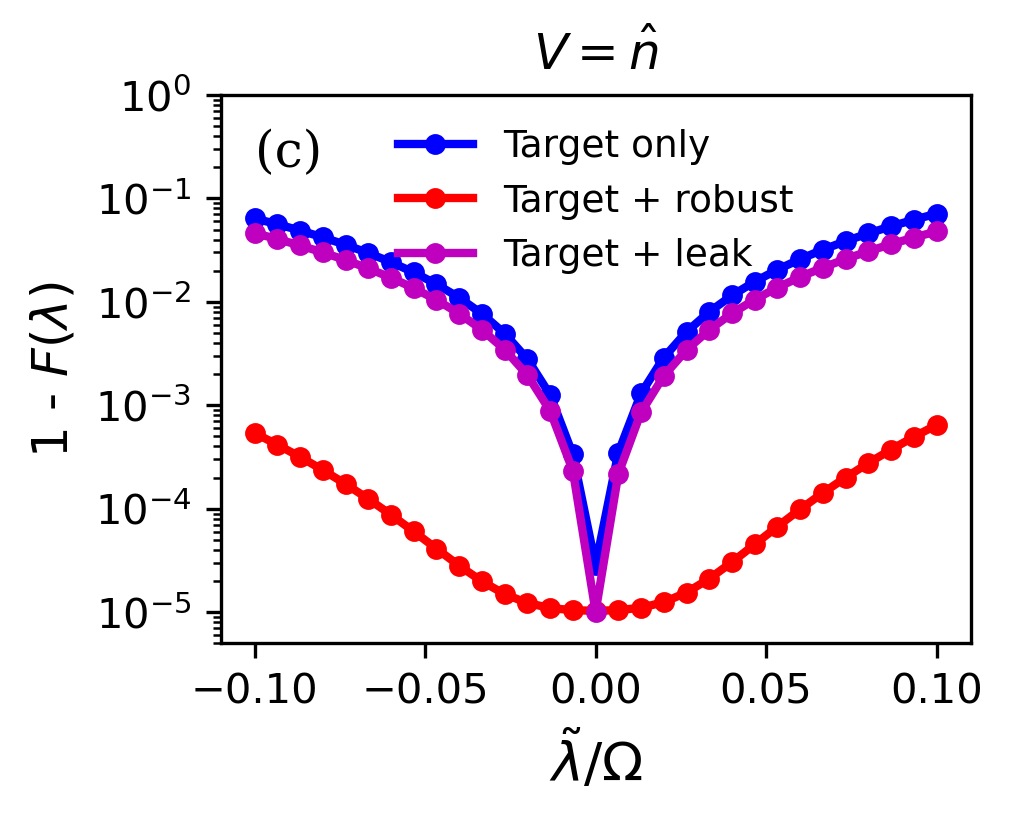}
    \caption{Comparison of the performance of the various optimal control protocols studied in the paper: target-only (blue), target \& robust (red), target \& leakage (purple), as well as DRAG (black dotted). Plots display (a) time-dependent target fidelity $f_0(t)$, (b) Time-dependent average leakage $l_0(t)$, and (c) final-time target infidelity $1-F_\lambda(t)$ in presence of a perturbation $\lambda V$. Results were obtained for $T/T_{\Omega}=1.3$, $\alpha/\Omega=-2$, and $V=\hat{n}$. Parameters for DRAG are chosen as in Fig. \ref{fig:perturb_plots}.   }
    \label{fig:dynamics_plots}
\end{figure*}

We now turn our attention to analyzing the role of leakage. For this we first study the performance of the two-stage optimization procedure where now the target cost $J_U$ (Eq. \ref{eq:cost_tar}) and the leakage cost $J_L$ (Eq. \ref{eq:cost_leak}) are minimized concurrently. We show results in Fig. \ref{fig:control_time_leak} in a regime similar to the target \& robustness optimization case of Figs. \ref{fig:control_time} and \ref{fig:control_time_alpha}. As in those cases, we observe a sharp threshold in the behavior of the optimized target functional $J_U$ and a smoother decay on the complementary functional, here $J_L$, as evolution time increases. In contrast, however, we see that increasing the magnitude of the anharmonicity $|\alpha|/\Omega$ does allow for further reducing leakage. We note also that the optimized leakage functional $J_L$ is non-monotonic with $T$. This can be understood from the fact that small values of $J_L$ can be either due to very short evolution times (where population has effectively no time to leak) or due to the control fields' ability to reduce leakage, which demands a finite control time. The results in Fig. \ref{fig:control_time_leak} indicate that, given enough evolution time, it is possible to achieve the desired target while reducing leakage outside the computational subspace at all times during the evolution. The leakage reduction improves as the anharmonicity $\alpha$ increases; this is reasonable as in the limit of $|\alpha|/\Omega\gg 1$ we expect the computational subspace of the oscillator to be effectively isolated from the rest of the levels. 

Up to this point, we have used the optimal control tools to derive three types of solutions: target-only (T) optimized, target \& robustness (T-R) optimized, and target \& leakage optimized (T-L). To get a deeper insight into how these solutions achieve the desired tasks, we compare in Fig. \ref{fig:dynamics_plots} (a) and (b) 
different dynamical properties of these protocols. In (a) we plot the time-dependent target-fidelity $f_0(t)$, which we obtain from evaluating Eq. (\ref{eq:fidP}) at $U_0(t)$
\begin{equation}
    f_0(t) = \frac{1}{6} \left\{\Tr[P U_0(t) P U_0(t)^\dagger]+ \left|\Tr[P U_0(t) P U_{\rm tar}^\dagger]\right|^2 \right\} , \label{eq:f0}
\end{equation}
Then, in (b) we plot the time-dependent leakage
\begin{equation}
    l_0(t)\equiv L[U_0(t)] = 1 - \frac{1}{2}\Tr\left(PU_0(t)PU_0(t)^\dagger\right),
\end{equation}
where $L(U)$ was defined in Eq. (\ref{eq:leakage_avg}). In this notation, the actual cost functions can be expressed as $J_{U} = 1-f_0(T)$ and $J_{L}=\int_0^T l_0(t) dt/T$. In addition, we include the evolution stemming from the DRAG protocol as a comparison. Note all these quantities are computed in absence of perturbation, and we take $T/T_{\Omega}=1.3$ and $\alpha/\Omega=-2$ for all protocols. 

Results in Fig. \ref{fig:dynamics_plots} confirm that T-L solutions are able to reach the target configuration while keeping leakage at a minimum ($1\%$ for the case shown) at all times. This is a factor of $10$ smaller than that shown by the T and T-R solutions, and is actually also smaller at times than the simple, first-order DRAG solution used for comparison. This confirms the effectiveness of the optimization procedure in the task of reducing loss out of the computational subspace throughout the dynamics, and not only at the final time. Upon adding the perturbation, however, we find that T-L solutions are as sensitive as regular target-only solutions. This is confirmed by the results in Fig. \ref{fig:dynamics_plots} (c), where we display infidelity at the final time in presence of a perturbation of strength $\lambda$, as in Fig. \ref{fig:perturb_plots}. 

These results hint at an apparent trade-off: we can improve the usual target-only optimal control to achieve robustness to static perturbations, but at the expense of having intermediate-time leakage. Conversely, we can augment the optimization procedure to reduce leakage at all times, but the dynamics remains sensitive to perturbations. This begs the question: can we obtain control solutions that achieve the target, are robust to perturbations, and also minimize leakage out of the computational subspace?.   

In order to explore this question, it would be useful to generalize the procedure outlined in Fig. \ref{fig:fig1}(b) to a higher number of cost functions. Here we take a simpler approach and retain the two-stage optimization procedure but now considering 
\begin{align}
    \rm{Stage\ A:} \  & J_A=J_U+J_R \\
    \rm{Stage\ B:}\  & J_B=J_L    
\end{align}

\begin{figure}
    \centering
    \includegraphics[width=\linewidth]{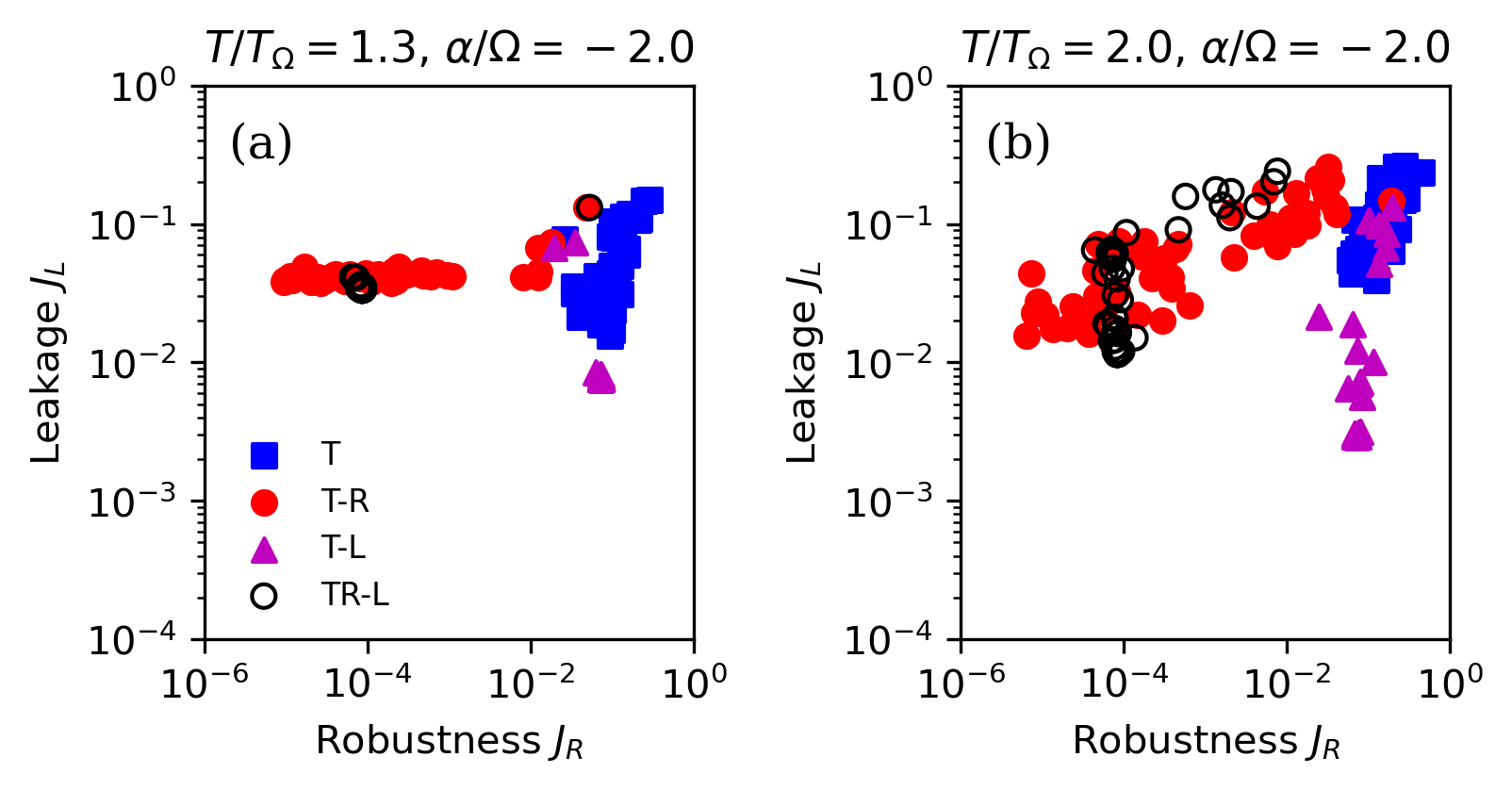}
    \includegraphics[width=\linewidth]{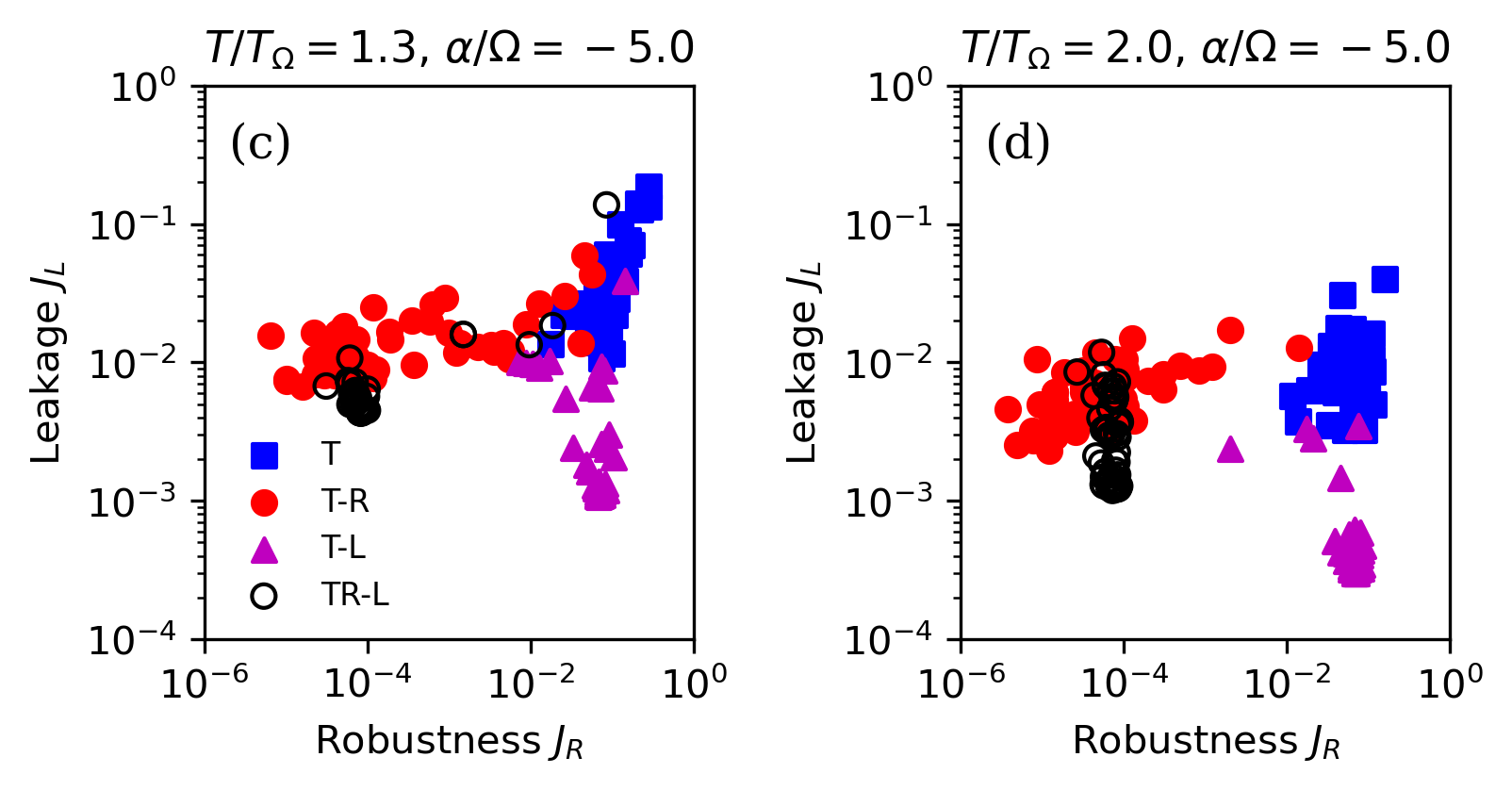}
    
    \caption{Robustness-leakage trade-off in the optimal control solutions. Scatter plots displaying solutions to various optimization procedures in a Robustness ($J_R$) - leakage ($J_L$) plane. Scenarios shown include optimization of target-only (T), target \& robustness (T-R), target \& leakage (T-L), and the target+robustness \& leakage (TR-L) case introduced in Sec. \ref{sec:leakage_tradeoff}.  
    }
    \label{fig:multi_opt}
\end{figure}

As discussed in Sec. \ref{sec:robust_qc}, choosing $J_A$ in this way guarantees that the global optimum is reached only when both terms $J_U$ and $J_R$ are separately minimized. In practice, we have observed that numerical values for near-optimal solutions of $J_U$ and $J_R$ are comparable (see e.g. Figs. \ref{fig:control_time} and \ref{fig:control_time_alpha}), and thus taking an equally-weighted sum is reasonable. We proceeded to run the two-stage optimization procedure in this new scenario, which we denote ``TR-L''. Interestingly, we found that the optimization struggles to find solutions which make all three functionals close to zero. Concretely, we observe no solutions which have $J_U+J_R\sim \varepsilon_A$ while concurrently reaching leakage values $J_L$ of the same order as those found in the target-leakage (non robust) optimization. To provide an overview of this behavior, we display in Fig. \ref{fig:multi_opt} scatter plots illustrating where different control solutions lay in the leakage-robustness plane. To do this we start with a random initial guess and run it through all four types of optimizations discussed in the paper: T (blue squares), T-R (red dots), T-L (purple triangles), and TR-L (black circles). The final value of both $J_R$ and $J_L$ is then recorded once the optimum for a given run has been reached. We then repeat this procedure for $M=50$ initializations. In Fig. \ref{fig:multi_opt} we display four scenarios corresponding to evolution times $T/T_\Omega \in \{ 1.3, 2.0\}$ and anharmonicities $\alpha/\Omega\in\{-2, -5\}$. These plots illustrate that T-R optimizations locate solutions of vanishing values of $J_R$ but always  displaying values $J_L$ which exceed the minimum achievable. The converse is true for the T-L optimization results, as purple triangles systematically cluster around $J_R \sim 10^{-1}$. Quite strikingly, we observe that the TR-L optimization, aimed at minimizing all three cost functions, is not able to significantly reduce leakage in the second optimization stage. We see that we when $|\alpha|$ increases (which effectively isolates the computational subspace in the system) a slight reduction of $J_L$ is achieved (as seen by comparing the location of  black circles with red dots in (c) and (d)). The reduction is more prominent if more evolution time is available. However, in the relevant case of finite $\alpha$ we observe an emergent trade-off where our procedure is not able to find solutions which are able to minimize leakage $J_L$ and sensitivity $J_R$ at the minimum possible levels. 

We conjecture this trade-off to be independent of our particular optimization framework. This is due to the fact that the incompatible costs $J_R$ and $J_L$ depend on the properties of the complete dynamics from $t=0$ to $t=T$, unlike $J_U$ which depends only on the final state at $t=T$. Schematically, this means that the T-R and T-L searches look for a path that minimizes a given property and ends in the desired point. However, a TR-L search demands looking for a single path that obeys two independent properties. Nonetheless, we leave a deeper numerical exploration using more sophisticated multi-objective optimization methods \cite{mavrotas2009} for future explorations.

An additional intuition for this conflict is as follows. In \cite{Poggi2024}, it was shown that universally robust control pulses require full controllability to dynamically generate a $1$-design \cite{gross2007} (a set of unitary matrices which mimic a uniform Haar distribution up to first-order moments) during the evolution in the full Hilbert space. However, many systems such as the transmon qubit are only fully controllable on the computational subspace and not on the full space. By constraining the evolution to a particular region of the Hilbert space, a $1$-design generating evolution cannot in general be obtained.

\section{Conclusions} \label{sec:conclusions}

We have studied the problem of optimal control of a quantum system where the target transformation lies in a computational subspace embedded in a larger Hilbert space. We propose two new cost functions that can be included in standard quantum optimal control methods, which allow for maximizing robustness to static perturbations and reducing leakage out of the computational subspace. We have derived optimal control protocols that either achieve a target gate while being robust, or achieve a target gate while minimizing leakage throughout the evolution. For these cases we have analyzed the resources required to implement these solutions in terms of anharmonicity and evolution time. Following this, we have identified an apparent trade-off between robustness and leakage, whereby standard optimization routines are not able to find control solutions that minimize both effects concurrently. 

In future work, we will consider the case of generalized robustness, where we seek to obtain a single control pulse that is robust to multiple orthogonal perturbation (e.g. $\hat{n}$ and $\hat{q}$), and explore how universal robustness (as defined in \cite{Poggi2024}) can be generalized to include only errors that affect the computational subspace.

\section*{Declaration of competing interest}
The authors declare that they have no known competing financial interests or personal relationships that could have appeared to
influence the work reported in this paper.

%%%%%%% Acknowledgments %%%%%%%

%\acknowledgments
\section*{Acknowledgments}
PMP acknowledges support by U.S. National Science Foundation (grant number PHY-2210013). AK acknowledges financial support of Taighde \'Eireann – Research Ireland under grant number 24/PATH-S/12701.

\appendix

\section{Fidelity susceptibility in a subspace \label{app:Fid_sus}}

As in \cite{Poggi2024}, we would like to quantify the robustness using the fidelity susceptibility i.e. the second derivative of the fidelity. We will assume that while population may leak outside the logical subspace during the ideal evolution, at the final time there is a block structure,
\begin{equation}
    U_0(T,0)= \left(\begin{matrix}
U_0^P & 0\\
0 & U_0^{\overline{P}}
\end{matrix}\right).
\end{equation}
It follows that $U_0(T,0)P U_0^\dagger(T,0)=P$.

For the purpose of calculation, we will make the following definitions,
\begin{eqnarray}
    A(\lambda) &=& \Tr[P U_\lambda P U_\lambda^\dagger] \\
    B(\lambda) &=& \left|\Tr[P U_\lambda P U_0^\dagger]\right|^2 \\
    &=& \left|\Tr[ P U_0^\dagger U_\lambda ]\right|^2 ,
\end{eqnarray}
since $P$ and $U_0(T,0)$ commute. We start by noting
\begin{eqnarray}
    A(0) &=& d_P ,\\
    B(0) &=& d_P^2 ,
\end{eqnarray}
which confirms the zeroth order terms as $F_0=1$.

Since $F_\lambda \leq 1$ and the maximum is at $\lambda=0$, we expect the first derivative to be zero. Let us confirm this now. Firstly we have,
\begin{eqnarray}
    A'(\lambda) &=& \Tr[P (\partial_\lambda U_\lambda) P U_\lambda^\dagger] +\Tr[P U_\lambda P (\partial_\lambda U_\lambda^\dagger)] \\
    &=& 2 \textrm{Re} \{ \Tr[P (\partial_\lambda U_\lambda) P U_\lambda^\dagger] \} \\
    &=& 2 \textrm{Re} \left\{-i \, \Tr\left[P \int_0^T ds U(T,s) V U_\lambda(s,0) P U_\lambda^\dagger\right] \right\}, \nonumber \\
\end{eqnarray}
which when evaluated at $\lambda=0$ gives
\begin{eqnarray}
    A'(0) &=& 2 \textrm{Re} \{-i T \Tr[U^\dagger_0 P U_0 \bar{V}_0 ] \} \\
    &=& 2 \textrm{Re} \{-i T \Tr[P \bar{V}_0 ] \} \\
    &=& 0.
\end{eqnarray}
\onecolumn
The other first derivative can be computed similarly as
\begin{eqnarray}
    B'(\lambda) &=& \frac{d}{d \lambda} \left| \Tr[P U_\lambda P U_0^\dagger]\right|^2 \\
    &=& 2 \textrm{Im} \left\{\Tr[P U_\lambda^\dagger U_0] \int_0^T ds \Tr\left[P U_0^\dagger U_\lambda(T,s) V U_\lambda(s,0)\right] \right\},
\end{eqnarray}
which when evaluated at $\lambda=0$ gives
\begin{eqnarray}
    B'(0) &=& 2 \textrm{Im} \left\{ \Tr(P) \int_0^T ds \Tr [P U_0^\dagger U_0 U^\dagger_0(s,0) V U_0(s,0)] \right\} \nonumber \\
    &=& 2 d_P T \textrm{Im}\left\{ \Tr[P \bar{V}_0]\right\} \\
    &=& 0.
\end{eqnarray}
Now that we have confirmed that the first derivative is indeed zero at $\lambda=0$, all that remains is to compute the second derivative. Before we begin, we note that the second derivative of the time evolution operator can be expressed as
\begin{eqnarray}
    \frac{d^2 U_\lambda}{d \lambda^2} = - U_0 \int_0^T ds \left[\int_s^T dx V(x) V(s) + \int_0^s dx V(s) V(x) \right],
\end{eqnarray}
where $V(t)\equiv U_0(t)^\dagger V U_0(t)$ and recall the following expressions for the first derivatives,
\begin{eqnarray}
    \frac{d U_\lambda}{d \lambda} \Big|_{\lambda=0} &=& -i U_0 \int_0^T ds V(s), \\
    \frac{d U_\lambda^\dagger}{d \lambda}\Big|_{\lambda=0} &=& i \int_0^T ds V(s) U_0^\dagger.
\end{eqnarray}
We start again with the first term,
\begin{eqnarray}
    A''(\lambda) &=& \Tr \left[P \frac{d^2 U_\lambda}{d \lambda^2} P U_\lambda^\dagger\right] + \Tr[P \frac{d U_\lambda}{d \lambda} P \frac{d U^\dagger_\lambda}{d \lambda}] + \textrm{c.c.} .
\end{eqnarray}
Let's evaluate this at $\lambda=0$ and apply the previous results, to give
\begin{eqnarray}
    A''(0) &=& - \Tr\left[P U_0^\dagger P U_0 \int_0^T ds \left(\int_s^T dx V(x)V(s)+\int_0^s dx V(s) V(x)\right)\right] + \Tr\left[P U_0 \int_0^T ds_1 V(s_1) P \int_0^T ds_2 V(s_2) U_0^\dagger\right] + \textrm{c.c.} \nonumber \\
    &=& 2 \Tr \left[ \left(\int_0^T ds P V(s)\right)^2\right]-\int_0^T ds \left[\int_s^T dx \Tr[P V(x)V(s)]+\int_0^s dx \Tr[V(x)V(s) P]\right] \nonumber \\
    &-&\int_0^T ds \left[\int_s^T dx \Tr[ V(s)V(x)P]+\int_0^s dx \Tr[P V(s)V(x)]\right] \\
    &=& 2 \Tr \left[\left(\int_0^T ds P V(s)\right)^2\right]-2 \Tr \int_0^T ds \int_0^T dx P V(x)V(s) \\
    &=& 2 \Tr \left[\left(P\int_0^T ds V(s)\right)^2\right] - 2 \Tr\left[P \left(\int_0^T ds V(s)\right)^2\right] \\
    &=& 2 T^2 \left\{\Tr[P \bar{V}_0 P \bar{V}_0] - \Tr[P \bar{V}_0^2]\right\} .
\end{eqnarray}
Finally, the second derivative of the second part is
\begin{eqnarray}
    B''(\lambda)&=& 2 \left|\int_0^T ds \, \Tr [P U_0^\dagger U_\lambda(T,s) V U_\lambda(s,0)]\right|^2 \nonumber \\ &-& 2 \textrm{Im} \left\{ i \Tr[U\lambda^\dagger U_0 P] \int_0^T ds \Tr \left[\int_s^T dx P U_0^\dagger U_\lambda (T,x) V U_\lambda(x,s) V U_\lambda(s,0)+\int_0^s dx P U_0^\dagger U_\lambda(T,s) V U_\lambda(s,x)V U_\lambda(x,0)\right]\right\}. \nonumber \\
\end{eqnarray}
At the relevant point $\lambda=0$, this simplifies dramatically,
\begin{eqnarray}
    B''(0) &=& 2 T^2 |\Tr_P \bar{V}_0|^2- 2 d_P \,\textrm{Im} \left\{i \int_0^T ds \left[\int_0^s dx \Tr[P V(s) V(x)]+ \int_s^T dx \Tr[P V(x)V(s)]\right]\right\} \\
    &=& 2 T^2 \left( |\Tr_P \bar{V}_0|^2 -d_P \Tr_P[\bar{V}_0^2] \right).
\end{eqnarray}
After all this calculation we arrive at our final result,
\begin{eqnarray}
    \frac{d^2 F_\lambda}{d \lambda^2} \Big|_{\lambda=0} = - \frac{2 T^2}{d_P} \left\{\Tr_P[\bar{V}_0^2]-\frac{1}{d_P+1}\left[ \Tr_P[\bar{V}_0]^2+\Tr_P[\bar{V}_0 P \bar{V}_0]\right] \right\},
\end{eqnarray}
\twocolumn
where we have defined $\Tr_P[\cdot]=\Tr[P \cdot]$. Note that in the limit $P \rightarrow 1$, this formula reduces to (up to some constant factors) the fidelity susceptibility derived in \cite{Poggi2024}.

As in the case of the formula for the full space, this susceptibility is completely minimised if $\bar{V}_0$ is proportional to the identity operator. If $\bar{V}_0$ is only the identity on the computational subspace, there are residual terms associated with the off diagonal block.

\section{DRAG pulses} \label{app:DRAG}

The first-order optimal DRAG pulses were derived in Ref. \cite{gambetta2011}. For the in-quadrature component, we choose a rescaled Gaussian function that starts and ends at zero, 
\begin{equation}
    d_R(t) = A \frac{\exp\left(-\frac{(t-T/2)^2}{2\sigma^2}\right) - \exp\left(-\frac{T^2}{8\sigma^2}\right)}{\sqrt{2\pi \sigma^2}\mathrm{erf}\left(\frac{T}{\sqrt{8\sigma^2}}\right) - T \exp\left(-\frac{T^2}{8\sigma^2}\right)  }.
\end{equation}

For the desired target gate, one chooses $\int_0^T d_R(t) dt=A=\pi$. Then, the other system parameters are chosen as

\begin{align}
    d_I(t) & = - \frac{\dot{d_R}(t)}{\sqrt{2}\alpha}, \\
    \delta(t) & = \frac{\Omega^2 d_R(t)^2}{2\alpha}(1-\sqrt{2}).
\end{align}

The parameters $\sigma$ and $T$ are chosen to optimize performance and robustness in the comparison shown in Sec. \ref{sec:results_robust}. 

\section{Scaling of leakage with anharmonicity} \label{app:Leak}

To get some intuition about the minimal operation time as a function of the anharmonicity, we consider a simple scenario. Consider implementing a $\pi$-pulse/ X gate with a real resonant field ($\delta=0$ and $d_I=0$), which furthermore is constant i.e. $d_R(t)=1$ for $t \in [0,T]$. The three level Hamiltonian in this case is
\begin{eqnarray}
    H_0(t) =\frac{1}{2} \left(
\begin{array}{ccc}
 0 & \Omega  & 0 \\
 \Omega  & 0 & \Omega  \sqrt{2} \\
 0 & \Omega  \sqrt{2} & 2 \alpha  \\
\end{array}
\right).
\end{eqnarray}
However for large values of $|\alpha|$, we can use adiabatic elimination to give an effective Hamiltonian,
\begin{eqnarray}
   H_{\rm eff}(t)= \frac{1}{2} \left(
\begin{array}{cc}
 0 & \Omega  \\
 \Omega  & -\frac{\Omega ^2}{\alpha } \\
\end{array}
\right).
\end{eqnarray}
Starting from the initial state $\ket{0}$, the probability to be in a state $\ket{1}$ at a final time $T$ is,
\begin{eqnarray}
    P_1 &=& \frac{4 \alpha ^2 \sin ^2\left(\frac{T \Omega  \sqrt{4 \alpha ^2+\Omega ^2}}{4 \alpha }\right)}{4 \alpha
   ^2+\Omega ^2} \\
   &\approx& \sin ^2\left(\frac{T \Omega }{2}\right) + \frac{T \Omega ^3 \sin (T \Omega )-4\Omega ^2 \sin ^2\left(\frac{T \Omega
   }{2}\right)}{16 \alpha ^2}, \nonumber \\
\end{eqnarray}
for large $|\alpha|$. For the standard pulse area of $\pi$ with $\Omega=\pi/T$, this simplifies to $P_1 = 1-\frac{\pi ^2}{4 \alpha ^2 T^2}$. Therefore from this heuristic argument, one would expect that to avoid the effects of leakage one would need suitably long operation time $T \gg 1/|\alpha|$.

%%%%%%%%  References  %%%%%%%%%
\bibliographystyle{elsarticle-num} 
\bibliography{transmon_refs}

\begin{thebibliography}{10}
\expandafter\ifx\csname url\endcsname\relax
  \def\url#1{\texttt{#1}}\fi
\expandafter\ifx\csname urlprefix\endcsname\relax\def\urlprefix{URL }\fi
\expandafter\ifx\csname href\endcsname\relax
  \def\href#1#2{#2} \def\path#1{#1}\fi

\bibitem{Wang2025_transmon}
Z.~Wang, E.~C. R.~W.~Parker, M.~S. Blok, Phys. Rev. Appl. 23 (2025) 034046.
\newblock
  \href{https://link.aps.org/doi/10.1103/PhysRevApplied.23.034046}{[link]}.
\newline\urlprefix\url{https://link.aps.org/doi/10.1103/PhysRevApplied.23.034046}

\bibitem{ruschhaupt2012}
A.~Ruschhaupt, X.~Chen, D.~Alonso, J.~Muga, New J. Phys. 14~(9) (2012) 093040.
\newblock \href{https://doi.org/10.1088/1367-2630/14/9/093040}{[link]}.
\newline\urlprefix\url{https://doi.org/10.1088/1367-2630/14/9/093040}

\bibitem{Poggi2024}
P.~M. Poggi, G.~De~Chiara, S.~Campbell, A.~Kiely, Phys. Rev. Lett. 132 (2024)
  193801.
\newblock
  \href{https://link.aps.org/doi/10.1103/PhysRevLett.132.193801}{[link]}.
\newline\urlprefix\url{https://link.aps.org/doi/10.1103/PhysRevLett.132.193801}

\bibitem{WEIDNER2025}
C.~A. Weidner, E.~A. Reed, J.~Monroe, B.~Sheller, S.~O’Neil, E.~Maas, E.~A.
  Jonckheere, F.~C. Langbein, S.~Schirmer, Automatica 172 (2025) 111987.
\newblock
  \href{https://www.sciencedirect.com/science/article/pii/S0005109824004813}{[link]}.
\newline\urlprefix\url{https://www.sciencedirect.com/science/article/pii/S0005109824004813}

\bibitem{bylander2011}
J.~Bylander, S.~Gustavsson, F.~Yan, F.~Yoshihara, K.~Harrabi, G.~Fitch, D.~G.
  Cory, Y.~Nakamura, J.-S. Tsai, W.~D. Oliver, Nat. Phys. 7~(7) (2011)
  565--570.
\newblock \href{https://doi.org/10.1038/nphys1994}{[link]}.
\newline\urlprefix\url{https://doi.org/10.1038/nphys1994}

\bibitem{guery2019}
D.~Gu{\'e}ry-Odelin, A.~Ruschhaupt, A.~Kiely, E.~Torrontegui,
  S.~Mart{\'\i}nez-Garaot, J.~G. Muga, Rev. Mod. Phys. 91~(4) (2019) 045001.
\newblock \href{https://doi.org/10.1103/RevModPhys.91.045001}{[link]}.
\newline\urlprefix\url{https://doi.org/10.1103/RevModPhys.91.045001}

\bibitem{kiely2014}
A.~Kiely, A.~Ruschhaupt, J. Phys. B 47~(11) (2014) 115501.
\newblock \href{https://doi.org/10.1088/0953-4075/47/11/115501}{[link]}.
\newline\urlprefix\url{https://doi.org/10.1088/0953-4075/47/11/115501}

\bibitem{motzoi2009_drag}
F.~Motzoi, J.~M. Gambetta, P.~Rebentrost, F.~K. Wilhelm, Phys. Rev. Lett.
  103~(11) (2009) 110501.
\newblock \href{https://doi.org/10.1103/PhysRevLett.103.110501}{[link]}.
\newline\urlprefix\url{https://doi.org/10.1103/PhysRevLett.103.110501}

\bibitem{babu2021}
A.~P. Babu, J.~Tuorila, T.~Ala-Nissila, npj Quantum Inf. 7~(1) (2021) 30.
\newblock \href{https://doi.org/10.1038/s41534-020-00357-z}{[link]}.
\newline\urlprefix\url{https://doi.org/10.1038/s41534-020-00357-z}

\bibitem{genov2013}
G.~T. Genov, N.~V. Vitanov, Phys. Rev. Lett. 110~(13) (2013) 133002.
\newblock \href{https://doi.org/10.1103/PhysRevLett.110.133002}{[link]}.
\newline\urlprefix\url{https://doi.org/10.1103/PhysRevLett.110.133002}

\bibitem{cykiert2024}
M.~Cykiert, E.~Ginossar, \href{https://doi.org/10.1088/1402-4896/ad7540}{Robust
  optimal control for a systematic error in the control amplitude of transmon
  qubits}, Phys. Scr. 99~(10) (2024) 105059.
\newline\urlprefix\url{https://doi.org/10.1088/1402-4896/ad7540}

\bibitem{tonchev2025}
H.~G. Tonchev, B.~T. Torosov, N.~V. Vitanov, Phys. Rev. A 112~(1) (2025)
  012605.
\newblock \href{https://doi.org/10.1103/s7h1-lqc1}{[link]}.
\newline\urlprefix\url{https://doi.org/10.1103/s7h1-lqc1}

\bibitem{rebentrost2009}
P.~Rebentrost, F.~K. Wilhelm, Phys. Rev. B 79~(6) (2009) 060507.
\newblock \href{https://doi.org/10.1103/PhysRevB.79.060507}{[link]}.
\newline\urlprefix\url{https://doi.org/10.1103/PhysRevB.79.060507}

\bibitem{safaei2009}
S.~Safaei, S.~Montangero, F.~Taddei, R.~Fazio, Phys. Rev. B 79~(6) (2009)
  064524.
\newblock \href{https://doi.org/10.1103/PhysRevB.79.064524}{[link]}.
\newline\urlprefix\url{https://doi.org/10.1103/PhysRevB.79.064524}

\bibitem{werninghaus2021}
M.~Werninghaus, D.~J. Egger, F.~Roy, S.~Machnes, F.~K. Wilhelm, S.~Filipp, npj
  Quantum Inf. 7~(1) (2021) 14.
\newblock \href{https://doi.org/10.1038/s41534-020-00346-2}{[link]}.
\newline\urlprefix\url{https://doi.org/10.1038/s41534-020-00346-2}

\bibitem{Hyyppa2024}
E.~Hyypp\"a, A.~Veps\"al\"ainen, M.~Papi\ifmmode~\check{c}\else \v{c}\fi{},
  C.~F. Chan, S.~Inel, A.~Landra, W.~Liu, J.~Luus, F.~Marxer,
  C.~Ockeloen-Korppi, S.~Orbell, B.~Tarasinski, J.~Heinsoo, PRX Quantum 5
  (2024) 030353.
\newblock \href{https://link.aps.org/doi/10.1103/PRXQuantum.5.030353}{[link]}.
\newline\urlprefix\url{https://link.aps.org/doi/10.1103/PRXQuantum.5.030353}

\bibitem{Dong2021}
W.~Dong, F.~Zhuang, S.~E. Economou, E.~Barnes, PRX Quantum 2 (2021) 030333.
\newblock \href{https://link.aps.org/doi/10.1103/PRXQuantum.2.030333}{[link]}.
\newline\urlprefix\url{https://link.aps.org/doi/10.1103/PRXQuantum.2.030333}

\bibitem{Coopmans2021}
L.~Coopmans, D.~Luo, G.~Kells, B.~K. Clark, J.~Carrasquilla, PRX Quantum 2
  (2021) 020332.
\newblock \href{https://link.aps.org/doi/10.1103/PRXQuantum.2.020332}{[link]}.
\newline\urlprefix\url{https://link.aps.org/doi/10.1103/PRXQuantum.2.020332}

\bibitem{Coopmans2022}
L.~Coopmans, S.~Campbell, G.~De~Chiara, A.~Kiely, Phys. Rev. Res. 4 (2022)
  043138.
\newblock
  \href{https://link.aps.org/doi/10.1103/PhysRevResearch.4.043138}{[link]}.
\newline\urlprefix\url{https://link.aps.org/doi/10.1103/PhysRevResearch.4.043138}

\bibitem{turyansky2025}
D.~Turyansky, Y.~Zolti, Y.~Cohen, A.~Pick, arXiv preprint arXiv:2507.09770
  (2025).
\newblock \href{https://doi.org/10.48550/arXiv.2507.09770}{[link]}.
\newline\urlprefix\url{https://doi.org/10.48550/arXiv.2507.09770}

\bibitem{Fresse-Colson2025}
O.~Fresse-Colson, S.~Gu\'erin, X.~Chen, D.~Sugny, Phys. Rev. A 112 (2025)
  022618.
\newblock \href{https://link.aps.org/doi/10.1103/rn5j-kjq3}{[link]}.
\newline\urlprefix\url{https://link.aps.org/doi/10.1103/rn5j-kjq3}

\bibitem{khodjasteh2012}
K.~Khodjasteh, H.~Bluhm, L.~Viola,
  \href{https://link.aps.org/doi/10.1103/PhysRevA.86.042329}{Automated
  synthesis of dynamically corrected quantum gates}, Phys. Rev. A 86~(4) (2012)
  042329.
\newline\urlprefix\url{https://link.aps.org/doi/10.1103/PhysRevA.86.042329}

\bibitem{Daems2013}
D.~Daems, A.~Ruschhaupt, D.~Sugny, S.~Gu\'erin, Phys. Rev. Lett. 111 (2013)
  050404.
\newblock
  \href{https://link.aps.org/doi/10.1103/PhysRevLett.111.050404}{[link]}.
\newline\urlprefix\url{https://link.aps.org/doi/10.1103/PhysRevLett.111.050404}

\bibitem{Araki2023}
T.~Araki, F.~Nori, C.~Gneiting, Phys. Rev. A 107 (2023) 032609.
\newblock \href{https://link.aps.org/doi/10.1103/PhysRevA.107.032609}{[link]}.
\newline\urlprefix\url{https://link.aps.org/doi/10.1103/PhysRevA.107.032609}

\bibitem{Irtaza2023}
I.~Khalid, C.~A. Weidner, E.~A. Jonckheere, S.~G. Shermer, F.~C. Langbein,
  Phys. Rev. A 107 (2023) 032606.
\newblock \href{https://link.aps.org/doi/10.1103/PhysRevA.107.032606}{[link]}.
\newline\urlprefix\url{https://link.aps.org/doi/10.1103/PhysRevA.107.032606}

\bibitem{Propson2022}
T.~Propson, B.~E. Jackson, J.~Koch, Z.~Manchester, D.~I. Schuster, Phys. Rev.
  Appl. 17 (2022) 014036.
\newblock
  \href{https://link.aps.org/doi/10.1103/PhysRevApplied.17.014036}{[link]}.
\newline\urlprefix\url{https://link.aps.org/doi/10.1103/PhysRevApplied.17.014036}

\bibitem{heeres2017}
R.~W. Heeres, P.~Reinhold, N.~Ofek, L.~Frunzio, L.~Jiang, M.~H. Devoret, R.~J.
  Schoelkopf, Nat. Commun. 8~(1) (2017) 94.
\newblock \href{https://doi.org/10.1038/s41467-017-00045-1}{[link]}.
\newline\urlprefix\url{https://doi.org/10.1038/s41467-017-00045-1}

\bibitem{abdelhafez2020}
M.~Abdelhafez, B.~Baker, A.~Gyenis, P.~Mundada, A.~A. Houck, D.~Schuster,
  J.~Koch, Phys. Rev. A 101~(2) (2020) 022321.
\newblock \href{https://doi.org/10.1103/PhysRevA.101.022321}{[link]}.
\newline\urlprefix\url{https://doi.org/10.1103/PhysRevA.101.022321}

\bibitem{gautier2025}
R.~Gautier, {\'E}.~Genois, A.~Blais, Phys. Rev. Lett. 134~(7) (2025) 070802.
\newblock \href{https://doi.org/10.1103/PhysRevLett.134.070802}{[link]}.
\newline\urlprefix\url{https://doi.org/10.1103/PhysRevLett.134.070802}

\bibitem{nguyen2024}
H.~N. Nguyen, F.~Motzoi, M.~Metcalf, K.~B. Whaley, M.~Bukov, M.~Schmitt, Mach.
  Learn.: Sci. Technol. 5~(2) (2024) 025066.
\newblock \href{https://doi.org/10.1088/2632-2153/ad4f4d}{[link]}.
\newline\urlprefix\url{https://doi.org/10.1088/2632-2153/ad4f4d}

\bibitem{papivc2023}
M.~Papi{\v{c}}, A.~Auer, I.~de~Vega, arXiv preprint arXiv:2305.08916 (2023).
\newblock \href{https://doi.org/10.48550/arXiv.2305.08916}{[link]}.
\newline\urlprefix\url{https://doi.org/10.48550/arXiv.2305.08916}

\bibitem{tripathi2024}
V.~Tripathi, H.~Chen, E.~Levenson-Falk, D.~A. Lidar, PRX Quantum 5~(1) (2024)
  010320.
\newblock \href{https://doi.org/10.1103/PRXQuantum.5.010320}{[link]}.
\newline\urlprefix\url{https://doi.org/10.1103/PRXQuantum.5.010320}

\bibitem{levy2018}
A.~Levy, A.~Kiely, J.~G. Muga, R.~Kosloff, E.~Torrontegui, New J. Phys. 20~(2)
  (2018) 025006.
\newblock \href{https://doi.org/10.1088/1367-2630/aaa9e5}{[link]}.
\newline\urlprefix\url{https://doi.org/10.1088/1367-2630/aaa9e5}

\bibitem{kosut2022}
R.~L. Kosut, G.~Bhole, H.~Rabitz, arXiv preprint arXiv:2208.14193 (2022).
\newblock \href{https://doi.org/10.48550/arXiv.2208.14193}{[link]}.
\newline\urlprefix\url{https://doi.org/10.48550/arXiv.2208.14193}

\bibitem{pedersen2007fidelity}
L.~H. Pedersen, N.~M. M{\o}ller, K.~M{\o}lmer, Phys. Lett. A 367~(1-2) (2007)
  47--51.
\newblock \href{https://doi.org/10.1016/j.physleta.2007.02.069}{[link]}.
\newline\urlprefix\url{https://doi.org/10.1016/j.physleta.2007.02.069}

\bibitem{fromonteil2023protocols}
C.~Fromonteil, D.~Bluvstein, H.~Pichler, PRX Quantum 4~(2) (2023) 020335.
\newblock \href{https://doi.org/10.1103/PRXQuantum.4.020335}{[link]}.
\newline\urlprefix\url{https://doi.org/10.1103/PRXQuantum.4.020335}

\bibitem{miettinen1999}
K.~Miettinen, \href{https://doi.org/10.1007/978-1-4615-5563-6}{Nonlinear
  multiobjective optimization}, Vol.~12, Springer Science \& Business Media,
  1999.
\newline\urlprefix\url{https://doi.org/10.1007/978-1-4615-5563-6}

\bibitem{shao2024}
B.~Shao, X.~Yang, R.~Liu, Y.~Zhai, D.~Lu, T.~Xin, J.~Li, Physical Review
  Applied 21~(3) (2024) 034042.
\newblock \href{https://doi.org/10.1103/PhysRevApplied.21.034042}{[link]}.
\newline\urlprefix\url{https://doi.org/10.1103/PhysRevApplied.21.034042}

\bibitem{kosut2023}
R.~L. Kosut, H.~Rabitz, in: 2023 IEEE International Conference on Quantum
  Computing and Engineering (QCE), Vol.~01, 2023, pp. 1304--1307.
\newblock \href{https://doi.org/10.1109/QCE57702.2023.00147}{[link]}.
\newline\urlprefix\url{https://doi.org/10.1109/QCE57702.2023.00147}

\bibitem{mavrotas2009}
G.~Mavrotas, Applied mathematics and computation 213~(2) (2009) 455--465.
\newblock \href{https://doi.org/10.1016/j.amc.2009.03.037}{[link]}.
\newline\urlprefix\url{https://doi.org/10.1016/j.amc.2009.03.037}

\bibitem{zhu1997}
C.~Zhu, R.~H. Byrd, P.~Lu, J.~Nocedal, ACM Trans. Math. Softw. 23~(4) (1997)
  550--560.
\newblock \href{https://doi.org/10.1145/279232.279236}{[link]}.
\newline\urlprefix\url{https://doi.org/10.1145/279232.279236}

\bibitem{byrd1995}
R.~H. Byrd, P.~Lu, J.~Nocedal, C.~Zhu, SIAM J. Sci. Comput. 16~(5) (1995)
  1190--1208.
\newblock \href{https://doi.org/10.1137/0916069}{[link]}.
\newline\urlprefix\url{https://doi.org/10.1137/0916069}

\bibitem{devoret2004}
M.~H. Devoret, A.~Wallraff, J.~M. Martinis, arXiv preprint cond-mat/0411174
  (2004).
\newblock \href{https://doi.org/10.48550/arXiv.cond-mat/0411174}{[link]}.
\newline\urlprefix\url{https://doi.org/10.48550/arXiv.cond-mat/0411174}

\bibitem{Gao2021}
Y.~Y. Gao, M.~A. Rol, S.~Touzard, C.~Wang, PRX Quantum 2 (2021) 040202.
\newblock \href{https://link.aps.org/doi/10.1103/PRXQuantum.2.040202}{[link]}.
\newline\urlprefix\url{https://link.aps.org/doi/10.1103/PRXQuantum.2.040202}

\bibitem{wang2025}
R.~Wang, Y.~Feng, Y.~Zhang, J.~Ding, B.~Li, F.~Motzoi, Y.~Gao, H.~Xu, Z.~Yang,
  W.~Nuerbolati, et~al., arXiv preprint arXiv:2502.10116 (2025).
\newblock \href{https://doi.org/10.48550/arXiv.2502.10116}{[link]}.
\newline\urlprefix\url{https://doi.org/10.48550/arXiv.2502.10116}

\bibitem{theis2018}
L.~Theis, F.~Motzoi, S.~Machnes, F.~Wilhelm, EPL 123~(6) (2018) 60001.
\newblock \href{https://doi.org/10.1209/0295-5075/123/60001}{[link]}.
\newline\urlprefix\url{https://doi.org/10.1209/0295-5075/123/60001}

\bibitem{gambetta2011}
J.~M. Gambetta, F.~Motzoi, S.~Merkel, F.~K. Wilhelm, Phys. Rev. A 83~(1) (2011)
  012308.
\newblock \href{https://doi.org/10.1103/PhysRevA.83.012308}{[link]}.
\newline\urlprefix\url{https://doi.org/10.1103/PhysRevA.83.012308}

\bibitem{deffner2017}
S.~Deffner, S.~Campbell, J. Phys. A 50~(45) (2017) 453001.
\newblock \href{https://doi.org/10.1088/1751-8121/aa86c6}{[link]}.
\newline\urlprefix\url{https://doi.org/10.1088/1751-8121/aa86c6}

\bibitem{poggi2019}
P.~M. Poggi, Phys. Rev. A 99~(4) (2019) 042116.
\newblock \href{https://doi.org/10.1103/PhysRevA.99.042116}{[link]}.
\newline\urlprefix\url{https://doi.org/10.1103/PhysRevA.99.042116}

\bibitem{duncan2025}
C.~W. Duncan, P.~M. Poggi, M.~Bukov, N.~T. Zinner, S.~Campbell, arXiv preprint
  arXiv:2501.16436 (2025).
\newblock \href{https://doi.org/10.48550/arXiv.2501.16436}{[link]}.
\newline\urlprefix\url{https://doi.org/10.48550/arXiv.2501.16436}

\bibitem{caneva2009}
T.~Caneva, M.~Murphy, T.~Calarco, R.~Fazio, S.~Montangero, V.~Giovannetti,
  G.~E. Santoro, Phys. Rev. Lett. 103~(24) (2009) 240501.
\newblock \href{https://doi.org/10.1103/PhysRevLett.103.240501}{[link]}.
\newline\urlprefix\url{https://doi.org/10.1103/PhysRevLett.103.240501}

\bibitem{bukov2018}
M.~Bukov, A.~G. Day, D.~Sels, P.~Weinberg, A.~Polkovnikov, P.~Mehta, Phys. Rev.
  X 8~(3) (2018) 031086.
\newblock \href{https://doi.org/10.1103/PhysRevX.8.031086}{[link]}.
\newline\urlprefix\url{https://doi.org/10.1103/PhysRevX.8.031086}

\bibitem{gross2007}
D.~Gross, K.~Audenaert, J.~Eisert, J. Math. Phys. 48~(5) (2007).
\newblock \href{https://doi.org/10.1063/1.2716992}{[link]}.
\newline\urlprefix\url{https://doi.org/10.1063/1.2716992}

\end{thebibliography}

\end{document}